\documentclass[10pt,twocolumn,twoside]{IEEEtran}

\usepackage[latin1]{inputenc}
\usepackage{color}
\usepackage{multirow,colortbl}
\usepackage[table]{xcolor}

\usepackage{cite}

\ifCLASSINFOpdf
  \usepackage[pdftex]{graphicx}
  \graphicspath{{Figures/},{./}}
  \DeclareGraphicsExtensions{.pdf,.jpeg,.png}
\else
\fi
\usepackage[cmex10]{amsmath}
\usepackage{amssymb}
\usepackage{array}

\usepackage{mdwmath}
\usepackage{fixltx2e}
\newcommand{\Z}{\mathbb Z}

\newcommand{\real}{{\mathbb R}}

\begin{document}

\title{Self-Similar Anisotropic Texture Analysis: the Hyperbolic Wavelet Transform Contribution}%

\author{S.G. Roux$^{(1)}$,   M. Clausel$^{(2)}$, B. Vedel$^{(3)}$,  S. Jaffard$^{(4)}$, P. Abry$^{(1)}$, IEEE Fellow\\
$\makebox{}$ \\
$^{(1)}$ Physics Dept., ENS Lyon, CNRS, UMR5672, Lyon, France.\\
{\tt stephane.roux@ens-lyon.fr, patrice.abry@ens-lyon.fr}\\
$^{(2)}$ Laboratoire Jean Kuntzmann, UMR 5224, University of Grenoble, France.\\
$^{(3)}$ LMBA, University of Bretagne Sud, European University of Bretagne, Vannes, France.\\
$^{(4)}$ University of Paris Est,  LAMA, CNRS, UMR 8050, Cr\'eteil, France.\\
}

\markboth{IEEE Trans. on Image Processing}%
{Hyperbolic Wavelet Transform for selfsimilar anisotropic image analysis}
%



\maketitle

\begin{abstract}
Textures in images can often be well modeled using self-similar processes while they may at the same time display anisotropy.
The present contribution thus aims at studying jointly selfsimilarity and anisotropy by focusing on a specific classical  class of Gaussian anisotropic selfsimilar processes.
It will first be shown that   accurate joint estimates of the anisotropy and selfsimilarity parameters  are performed by  replacing the standard 2D-discrete wavelet transform by the hyperbolic wavelet transform, which permits the use of different dilation factors along the horizontal and vertical axis.
Defining anisotropy requires a reference direction that needs not a priori match the horizontal and vertical axes according to which the images are digitized, this discrepancy defines a rotation angle.
Second, we show that  this rotation angle can be jointly estimated.
Third, a non parametric bootstrap based procedure is described, that provides confidence interval in addition to the estimates themselves and enables to construct an isotropy test procedure, that can be applied to a single texture image.
Fourth, the robustness and versatility of the proposed analysis is illustrated by being applied to a large variety of different isotropic and anisotropic self-similar fields.
As an illustration, we show that  a true anisotropy built-in self-similarity can be disentangled from an isotropic self-similarity to which an anisotropic trend has been superimposed.
\end{abstract}

%


\section{Introduction}
\label{sec:intro}

\noindent {\bf Texture classification.} \quad In numerous modern applications (satellite imagery \cite{RAD2000}, geography \cite{f97}, biomedical imagery \cite{Lundahl:1986,aK2001,Rachidi::2008,Richard::2010,B2010}, geophysics \cite{SL87}, art investigation \cite{ABRY:2012:A}, \ldots), the data available for analysis consist of images of homogeneous textures, that need to be characterized.
Texture classification thus consists of classical problem in image processing that received considerable efforts in recent years (cf. e.g.,  \cite{Lundahl:1986,U1995,N02,D2002,C05,vsv99,P84,K99} and references therein). \\*[-.2cm]

\noindent {\bf Scale invariance and Self-similarity.} \quad Amongst the many different ways texture characterizations have been investigated, techniques based on  \emph{scale invariance}, or \emph{fractal}, concepts are considered as promising, notably for the application fields listed above (cf. e.g., \cite{Pesquet-Popescu::2002} for a review).
Scale invariance can be defined as the fact that there exists no specific space-scale in data that play a preferred role in their space dynamic, or equivalently that all space-scales are equally important.
Scale invariance in data implies that they are analyzed with (statistical) models that do not rely on the identification of specific scales (such as Markov models) but instead with models that aim at characterizing a relation amongst scales.
Because \emph{Self-Similarity} is a theoretically well-grounded, and relatively simple instance of scale invariance behaviors, it has often been proposed that Gaussian self-similar fields are relevant models enabling efficient characterization and classification of the textures (cf. e.g, \cite{K99,P84}). \\*[-.2cm]

\noindent {\bf Anisotropy.} \quad However, textures are also often characterized by anisotropy, which may either be deeply tied to self-similarity itself \cite{BE03,BMS07} or exist as an independent property that is superimposed to an isotropic self-similarity.
In both cases, it is a crucial stake in analysis to disentangle self-similarity from anisotropy, to discriminate whether self-similarity and anisotropy are independent properties or if they are stemming from the same constructive mechanism, as well as to be able to estimate accurately the self-similarity parameter $H$, despite anisotropy.
It has already been pointed out that fractal analysis and estimation is very sensitive to anisotropy (cf. e.g., \cite{SL87}).
In the literature, anisotropy is often analyzed from 1D slices extracted from images along different directions \cite{Richard::2010} or by making use of local directional differential estimators {\cite{R92,S94}.  \\*[-.2cm]

\noindent {\bf Goals and contributions.} \quad In this context, elaborating on a preliminary attempt \cite{rcvja11}, the present contribution aims at proposing an efficient and elegant solution to the joint analysis and estimation of self-similarity and anisotropy in 2D fields. 
Though the devised procedure aims at, and is designed for, being applied to real-world textures, its performance are assessed by means of Monte Carlo simulation performed on synthetic isotropic and non isotropic Gaussian textures.
While the (discrete) wavelet transform (DWT)  is nowadays a classical tool for image processing, the key originality of the present work is to show that the classical discrete wavelet transform fails at providing a relevant analysis of self-similarity in presence of anisotropy and it is instead here proposed to replace it with the \emph{Hyperbolic Wavelet Transform} (HWT) (defined e.g.,  in \cite{DeVore1998}).
Indeed, the use of different dilation factors on the axes $x$ and $y$ potentially permits to \emph{see} the anisotropy, as opposed to the classical 2D Discrete Wavelet Transform relying on a single and isotropic dilation operator. 
The Hyperbolic Wavelet Transform is defined in Section~\ref{sec:HWT}. 
Note that the HWT had appear earlier in the literature under different names, such as \emph{Tensor-product wavelet} \cite{Yu96}, \emph{anisotropic wavelet transform} \cite{Ro1999} or \emph{rectangular wavelet transform} \cite{Zav07}, without specific exploration though of its benefits to study anisotropy in textures. 
Also, redundant (or overcomplete) wavelet representations (such as M-band, dual tree and Hilbert pair complex wavelets, cf. e.g., \cite{S05} for a enlightening review) may be used to analyze images and textures. 
However, while they suffer from a larger computational cost, they have been observed (in preliminary attempts performed by the authors) to yield little, if not no, practical benefit for the study of scale invariance and the estimation of the corresponding parameters. 
Overcomplete wavelet representations are thus excluded from the present study. 

As representative of 2D model mixing self-similarity and anisotropy, self-similar Gaussian 2D random fields with built-in anisotropy, such as those proposed in e.g., \cite{BE03,BMS07}, are used here.
These processes are defined and illustrated in Section~\ref{sec:OSGRF} and their hyperbolic wavelet analysis is detailed in Section~\ref{sec:HWTOSGRF}.

Estimation procedures for the parameters characterizing self-similarity and anisotropy are defined and their performance assessed in Section~\ref{sec:esta}.
The definition of anisotropy involves a rigid definition of reference (orthogonal) axes, that have no reason a priori to match those of the sensor used to acquire the image and thus to coincide with the horizontal $x$ and vertical $y$ axis with which the image is presented for analysis.
Therefore, the model introduced in Section~\ref{sec:OSGRF} includes a rotation parameter that accounts for this unknown.
An estimation procedure for this rotation is devised and analyzed in Section~\ref{sec:estb}.
Therefore, the parameters characterizing rotation, anisotropy and self-similarity are estimated jointly.

For application purposes, it is crucial to be able to decide whether textures should be modeled by isotropic or anisotropic models.
Therefore, a procedure for testing the null hypothesis that the texture is isotropic is constructed and studied in Section~\ref{sec:test}.
It is based on a non parametric bootstrap procedure performed on the hyperbolic wavelet coefficients (in the spirit of the construction devised in \cite{WENDT:2007:E}) and can thus be applied to each single analyzed texture independently.
Incidentally, the bootstrap procedures also provides us with confidence intervals for the estimates, a very important feature for practical purposes.

To finish with, the analysis procedures proposed here are applied in Section~\ref{sec:othermodel} to a variety of isotropic and anisotropic fields that differ from the precise model used as a reference model (cf. Section~\ref{sec:OSGRF}), hence illustrating the robustness and generality of the tools proposed here.
Notably, it is shown that the proposed analysis enables to clearly distinguish between a truly anisotropic self-similar field from a isotropic self-similar field (with same self-similar parameter) to which directional, hence anisotropic, oscillations have been additively superimposed.
Joint anisotropy and self-similarity is hence clearly disentangled from isotropic self-similarity, with unrelated superimposed anisotropic trend.

\begin{figure}[t]
\begin{minipage}[b]{1.0\linewidth}
\includegraphics[width=.33\linewidth]{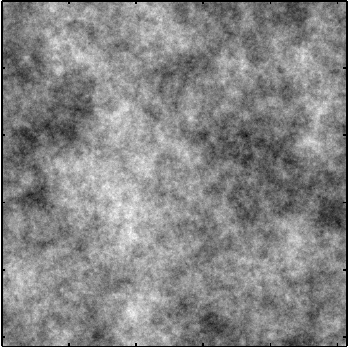}~\includegraphics[width=.33\linewidth]{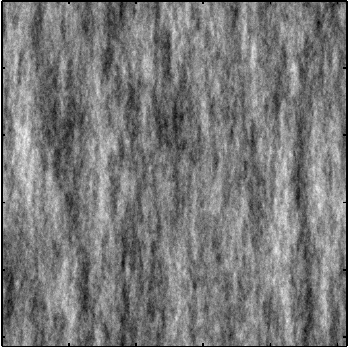}~\includegraphics[width=.33\linewidth]{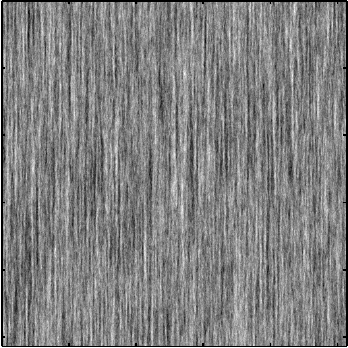}\\*[.04cm]
\includegraphics[width=.33\linewidth]{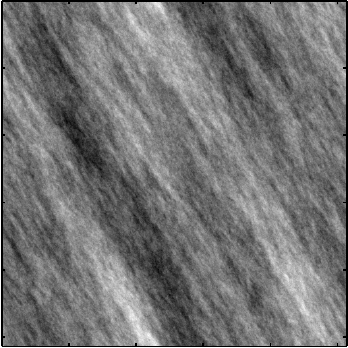}~\includegraphics[width=.33\linewidth]{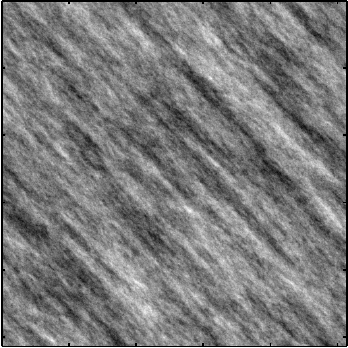}~\includegraphics[width=.33\linewidth]{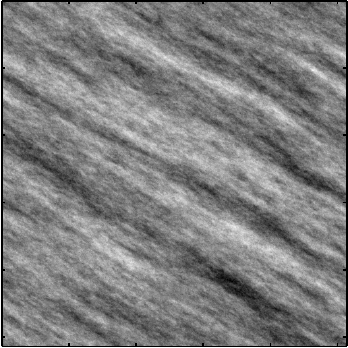}
\end{minipage}
\vspace{-.2cm}
\caption{\label{fig:model} {\bf Sample fields of $X_{\theta_0,\alpha_0,H_0}$.} Top line : $(\theta_0,\alpha_0,H_0)=(0,\alpha_0,0.2)$ with, from left to right $\alpha_0=1$ (isotropic); $\alpha_0=0.7$ and $\alpha_0=0.3$. Bottom line : $(\theta_0,\alpha_0,H_0)=(\theta_0,0.7,0.2)$ and, from left to right $\theta_0=\pi/6$, $\theta_0=\pi/4$ and $\theta_0=\pi/3$.}
\end{figure}

\section{Hyperbolic Wavelet Analysis of anisotropic self-similar random fields}

\subsection{Anisotropic self-similar random fields}
\label{sec:OSGRF}

\subsubsection{Definition}

Because of its generic and representative nature, it has been chosen to work with the class of anisotropic Gaussian self-similar fields, introduced in \cite{BE03,BMS07},  referred to as \emph{Operator Scaling Gaussian Random Field} ({\bf OSGRF}) {which can be defined using the following harmonizable representation}:
\begin{equation}
{X_{f,E_0,H_0}(\underline{x})} =  \int_{\mathbb{R}^2} (e^{i\langle \underline{x}, \, \underline{\xi} \rangle} -1) f(\underline{\xi})^{-(H_0 +1)} d\widehat{W}(\underline{\xi})\,,
\label{eq:modela}
\end{equation}
where $\underline{x}=(x_1,x_2)$,  $\underline{\xi}=(\xi_1,\xi_2)$, $E_0$ is a matrix satisfying Tr$(E_0)=2$,
$f$ a  { $E_0$--homogeneous continuous positive function} (hence   {satisfying the homogeneity relationship
$f(a^{E_0}\underline{\xi})=af(\underline{\xi})$ on $\mathbb{R}^2$)
such that $\int (1\wedge |\underline{\xi}|^2)f(\underline{\xi})^{-2(H_0 +1)} d\underline{\xi}<+\infty$, and  where $d\widehat{W}(\underline{\xi})$ stands for a 2D  {Wiener measure}.}

When $f$ is not a radial function, the Gaussian field is not isotropic.
In this study, it is chosen to use the following 2-parameter (related to anisotropy and rotation) explicit form:
$$ f_{\theta_0,\alpha_0}(\underline{\xi})=(\vert \zeta_1 \vert^{1/\alpha_0} + \vert \zeta_2 \vert^ {1/(2-\alpha_0)}),$$
with $\underline{\zeta}=(\zeta_1,\zeta_2)=R_{\theta_0} \underline{\xi}$ and rotation matrix $\displaystyle  R_{\theta_0}$ defined as:
$$\displaystyle  R_{\theta_0} = \begin{pmatrix}\cos(\theta_0)&-\sin(\theta_0)\\\sin(\theta_0) & \cos(\theta_0)\end{pmatrix}.$$
In this model,  $E_0=\begin{pmatrix}\alpha_0&0\\0&2-\alpha_0\end{pmatrix}$,  $0<\alpha_0<2$.

A rotation parameter $\theta_0$ has been added to  the definition \ref{eq:modela} : It accounts for the fact that the rigid axes according to which anisotropy is defined need not match a priori the sensor axes according to which the image is digitalized  (for real-world data) or numerically produced (for synthetic textures).
In the sequel, {\bf OSGRF} $X_{\theta_0,\alpha_0,H_0}$ thus refers to the following model, relying on $3$ parameters, $\theta_0,\alpha_0,H_0$, characterizing respectively, rotation, anisotropy and self-similarity
\begin{eqnarray}
X_{\theta_0,\alpha_0,H_0}(\underline{x}) = &  \int_{\mathbb{R}^2} (e^{i\langle \underline{x}, \, \underline{\xi} \rangle} -1) f_{\alpha_0,\theta_0}(\underline{\xi})^{-(H_0 +1)} d\widehat{W}(\underline{\xi})
\label{eq:model}
\end{eqnarray}

\subsubsection{Properties}
With this construction, {\bf OSGRF} $X_{\theta_0,\alpha_0,H_0}$ has stationary increments.
It possesses a built-in anisotropy characterized by the parameter $ \alpha_0 \in (0, 2)$.
When $\alpha_0=1$, the field is isotropic and the case $1<\alpha_0<2$ correspond to the case $0<\alpha_0<1$ with the axes $(x_1,x_2)$ permuted.

{\bf OSGRF} $X_{\theta_0,\alpha_0,H_0}$
satisfies  (where $\overset{\mathcal{L}}{=}  $ denotes equality for all finite dimensional distributions):
\begin{equation}\label{e:selfsimilar}
\{ X_{\alpha_0,H_0,\theta_0}(a^{E}  \underline{x} )\}
\overset{\mathcal{L}}{=} \{ a^{H_0} X_{\alpha_0,H_0,\theta_0}( \underline{x}) \}.
\end{equation}
with $E_0=R_{\theta_0}E R_{-\theta_0}$. It is thus exactly self-similar with parameter $0< H_0 < \makebox{ min }(\alpha_0, 2-\alpha_0) <2 $.

Fig.~\ref{fig:model} displays realizations of $X_{\theta_0,\alpha_0,H_0}(\underline{x})$, obtained from {\sc Matlab} routines written by ourselves and available upon request.
On top row, $(H_0,\theta_0)=(0.2,0)$ are kept fixed while $\alpha_0$ is varied from  $\alpha_0 =1$, $0.7$ and $0.3$  (from left to right).
The practical goal is to estimate correctly $H_0$ despite these different unknown anisotropy strengths $\alpha_0$.
On bottom row, a strongly anisotropic field is shown,  $(\alpha_0=0.3,H_0=0.2)$, with $3$ different rotation angles $\theta_0$
(from left to right  $\theta_0=\pi/6$, $\pi/4$ and $\pi/3$).
The targeted goal is here to estimate correctly $(\alpha_0,H_0)$ despite such unknown rotation.

This three-parameter {\bf OSGRF} stochastic process  provides us with a rich and versatile model for selfsimilar (an)isotropic textures.

\subsubsection{Numerical simulation}
Realizations (or sample fields) of the synthetic processes defined in Eq.~(\ref{eq:model}) are produced numerically following the classical procedure, recalled in e.g., \cite{BMS07}, relying on drawing at random realizations of white-noise $d\widehat{W}(\underline{\xi})$, followed by standard numerical integration procedures.  


\begin{figure}[t]
\begin{minipage}[b]{1.0\linewidth}
  \centering
  \centerline{\includegraphics[width=1.0\linewidth]{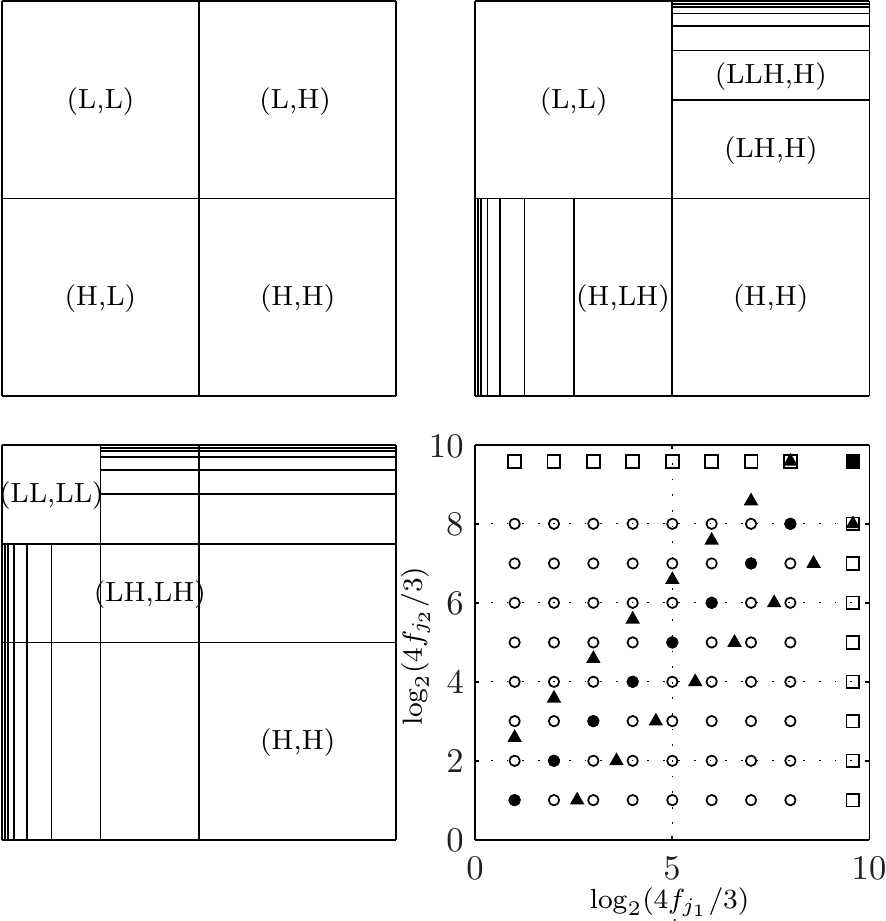}}
\end{minipage}
\vspace{-.6cm}
\caption{\label{fig:algo} {\bf Hyperbolic Wavelet Transform.}
Top line : one step of HWT consists of one step of 2D-DWT (left) (with 1D-DWT performed on each line of HL and each column of LH  subbands (right)).
Bottom line : second step of HWT (left) and locations of the HWT vs DWT (black dots) coefficients in the frequency domain (right).
Black dots correspond to 2D-DWT while for HWT, white dots indicate subband HH, black triangles indicate LH and HL, squares correspond to the approximation coefficients).
The circle symbols correspond to the coefficients $\psi_{j_1,j_2}$ $j_1,j_2\neq 0$  and square symbols  to  the coefficients $\psi_{0,j_2}, \psi_{j_1,0}, \psi_{0,0}$.
}
\end{figure}

\subsection{Hyperbolic Wavelet Transform}
\label{sec:HWT}

The 2D Hyperbolic Wavelet Transform (HWT) differs from the 2D Discrete Wavelet Transform (DWT) insofar as its definition relies on the use of two different dilation factors along the horizontal and vertical axes, as opposed to the 2D-DWT that makes use of a single and same dilation factor along both axes.
This difference turns out to be crucial for the analysis of anisotropy.

The collection of functions constituting the orthogonal basis underlying the HWT is defined as tensor products of univariate wavelets (cf. e.g., \cite{Mallat1998}).
Let $\varphi$ and $\psi$ denote the scaling function and the wavelet of a given one-dimensional multiresolution analysis.  
The HWT basis of ${\cal L}^2(\real^2)$ is defined as (cf.~\cite{DeVore1998}):
\begin{eqnarray*}
\psi_{j_1,j_2,k_1,k_2}(x_1,x_2)\!\!\!\!\!&=&\!\!\!\!\!
\psi(2^{j_1}x_1-k_1)\psi(2^{j_2}x_2-k_2),\\
\psi_{0,j_2,k_1,k_2}(x_1,x_2)\!\!\!\!\!&=&\!\!\!\!\!
\varphi(x_1-k_1)\psi(2^{j_2}x_2-k_2),\\
\psi_{j_1,0,k_1,k_2}(x_1,x_2)\!\!\!\!\!&=&\!\!\!\!\!
\psi(2^{j_1}x_1-k_1)\varphi(x_2-k_2),\\
\psi_{0,0,k_1,k_2}(x_1,\!\!\!\!\!&x_2&\!\!\!\!\!)= \varphi(x_1-k_1)\varphi(x_2-k_2),
\end{eqnarray*}
for all  $\underline{j}=(j_1,j_2)\in \mathbb{N}^{*2}$ and $\underline{k}=(k_1,k_2) \in \Z^2$.
The HWT shares a deep relation with Triebel bases, used in mathematic literature, to characterize anisotropic functional spaces (cf.~\cite{Trie06}).

The hyperbolic wavelet coefficients of the process $X$
are defined, $\forall \underline{j}=(j_1,j_2)\in\mathbb{N}^{*2}$,  as :
\begin{equation}
d_X(\underline{j},\underline{k})= 2^{j_1+j_2}\int_{\mathbb{R}^2}\psi_{\underline{j},\underline{k}}(x_1,x_2)X(x_1,x_2)d
x_1 d x_2.
\label{eq:coef}
\end{equation}
Note that a ${\cal L}^1$-normalization is used (instead of the classical ${\cal L}^2$-norm), as it better suits self-similarity analysis (cf. e.g., \cite{BMA93}).
With these notations, fine resolution scales correspond to the limit $2^{j_1}, 2^{j_2} \rightarrow  +\infty$, and index $j$ in the decomposition corresponds to the actual resolution $2^{J-j}$, where $J=\log_2(N)$, for an image of size ($N\times N$).

Such coefficients $d_X(\underline{j},\underline{k})$ can be computed efficiently, using a recursive pyramidal filter bank based algorithm comparable to that underlying the 2D-DWT.
In Fig.~\ref{fig:algo}, the first  two iterations are illustrated, in the Fourier domain.
One iteration of HWT practically consists of  the combination of one iteration of the
2D-DWT algorithm, with 1D-DWT  performed on each line of the vertical details (HL) and 1D DWT performed on each column of the horizontal details (LH).
Because the central frequencies of the dilated scaling function of $\varphi(2^j x)$ and mother-wavelet $\psi(2^j x)$ can be approximated as $f_j=\frac{1}{4}2^{j}$ and $f_j=\frac{3}{4}2^{j}$, respectively, the HWT coefficients $d_X(\underline{j},\underline{k})$ can be located in a (log-) frequency-frequency plane as shown in   Fig.~\ref{fig:algo} bottom right) and thus compared to the location of the 2D-DWT coefficients.

In what follows, a 1D-Daubechies-$3$ multiresolution is used \cite{daubechies92}.

\subsection{Analysis of anisotropic self-similar random fields}
\label{sec:HWTOSGRF}

From Eq. (\ref{eq:coef}),  the HWT coefficients  can be rewritten, $\forall \underline{j}=(j_1,j_2)\in\mathbb{N}^{*2}$ and
$\forall \underline{k}=(k_1,k_2)\in\mathbb{Z}^2$,  as stochastic integrals:
\begin{equation}
\label{def-wc}
d_X(\underline{j},\underline{k})=\int_{\mathbb{R}^2}
\frac{\left(\prod_{\ell=1}^{2}e^{i
2^{-j_\ell}k_\ell\xi_\ell}\overline{\widehat{\psi}(2^{-j_\ell}\xi_\ell)}\right)}
{\left(|\xi_1|^{1/\alpha_0}+|\xi_2|^{1/(2-\alpha_0)}\right)^{H_0+1}}\;d\widehat{W}(\xi_1,\xi_2)
\;.
\end{equation}
{Following the methodology in ~\cite{CV10b,ACJRV2012}, it can be proven that the HWT coefficients are weakly correlated,  i.e.,  $\forall (j_1,j_2,k_1,k_2,k'_1,k'_2)$
\begin{multline}
\left|\mathbb{E}(d_X((j_1,j_2),(k_1,k_2))d_X((j_1,j_2),(k'_1,k'_2)))\right| \\
\leq
\frac{\mathbb{E}(|d_X((j_1,j_2,0,0)|^2)}{1+|k_1-k'_1|+|k_2-k'_2|}\;.
\end{multline}
Using the substitution
$\zeta_1=2^{-j_1}\xi_1$,
$\zeta_2=2^{-j_2}\xi_2$ in the rewriting of definition of the wavelet coefficients (cf. Eq.~(\ref{def-wc})), we have been able to show that  the HWT coefficients typically behave as~\cite{CV10b,ACJRV2012} :
\[
d_X(\underline{j},\underline{k})\simeq 2^{\frac{j_1+j_2}{2}}\int_{\mathbb{R}^2}
\frac{\left(\prod_{\ell=1}^{2}e^{i
k_\ell\zeta_\ell}\overline{\widehat{\psi}(\zeta_\ell)}\right)d\widehat{W}(\zeta_1,\zeta_2)}
{\left(2^{\frac{j_1}{\alpha_0}}|\zeta_1|^{\frac{1}{\alpha_0}}+2^{\frac{j_2}{2-\alpha_0}}|\zeta_2|^{\frac{1}{2-\alpha_0}}\right)^{H_0+1}}
\;,
\]
When $j_1/\alpha_0>j_2/(2-\alpha_0)$, we derive that, for all $\zeta_1,\zeta_2$:
\begin{multline*}
2^{\frac{j_1}{\alpha_0}}|\zeta_1|^{\frac{1}{\alpha_0}} \leq 
2^{\frac{j_1}{\alpha_0}}|\zeta_1|^{\frac{1}{\alpha_0}}+2^{\frac{j_2}{2-\alpha_0}}|\zeta_2|^{\frac{1}{2-\alpha_0}} \leq \\
 2^{\frac{j_1}{\alpha_0}}\left(|\zeta_1|^{\frac{1}{\alpha_0}}+|\zeta_2|^{\frac{1}{2-\alpha_0}}\right)
\end{multline*}
and further 
\begin{multline*}
2^{\frac{-j_1H_0}{\alpha_0}}
\frac{1}{(|\zeta_1|^{\frac{1}{\alpha_0}}+|\zeta_2|^{\frac{1}{2-\alpha_0}})^{H_0+1}} \leq  \\
\frac{1}{2^{\frac{j_1}{\alpha_0}}|\zeta_1|^{\frac{1}{\alpha_0}}+2^{\frac{j_2}{2-\alpha_0}}|\zeta_2|^{\frac{1}{2-\alpha_0}}}\leq
2^{\frac{-j_1H_0}{\alpha_0}}\frac{1}{\left(|\zeta_1|^{\frac{H_0+1}{\alpha_0}}\right)}
\end{multline*}
which enabled us to obtain the following inequality:
\begin{multline*}
C_1 2^{\frac{j_1+j_2}{2}}2^{-\frac{j_1(H_0+1)}{\alpha_0}}\leq
\mathbb{E}(|d_X(\underline{j},\underline{k})|^2)^{1/2}\leq \\ C_2
2^{\frac{j_1+j_2}{2}}2^{-\frac{j_1(H_0+1)}{\alpha_0}}\; ,
\end{multline*}
\[
\makebox{ with } C_1=\left(\int_{\mathbb{R}^2}
\frac{\prod_{\ell=1}^{2}|\widehat{\psi}(\zeta_\ell)|^2d\xi}{(|\zeta_1|^{\frac{1}{\alpha_0}}+|\zeta_2|^{\frac{1}{2-\alpha_0}})^{H_0+1}}\right)^{1/2}\;,
\]
\[
\makebox{ and } C_2=\left(\int_{\mathbb{R}^2} 	
\frac{\prod_{\ell=1}^{2}|\widehat{\psi}(\zeta_\ell)|^2 d\xi}{(|\zeta_1|^{\frac{H_0+1}{\alpha_0}})}\right)^{1/2}\;.
\]
}
Combined to similar arguments for the case $j_1/\alpha_0\leq j_2/(2-\alpha_0)$, these calculations enable us to show that the order of magnitude of the expectation of the squared HWT coefficients reads, $\forall (j_1,j_2)$:
\begin{equation}
\label{eq:HWT}
\mathbb{E}(|d_X(\underline{j},\underline{k})|^2)^{1/2}\approx
2^{\frac{j_1+j_2}{2}}2^{-(H_0+1)\max(\frac{j_1}{\alpha_0},\frac{j_2}{2-\alpha_0})}\;.
\end{equation}
Because the processes of interest here are Gaussian, this can straightforwardly be extended to any  $ q \geq  -2 $, cf. \cite{CV10a,CV10b,ACJRV2012}: 
\begin{equation}
\label{eq:HWTq}
\mathbb{E}(|d_X(\underline{j},\underline{k})|^q)\approx
2^{q(j_1+j_2)}2^{-q(H_0+1)\max(\frac{j_1}{\alpha_0},\frac{j_2}{2-\alpha_0})}\;.
\end{equation}
These key results constitute the founding ingredient for the estimation procedures defined below. 

\section{Parameter Estimation}
\label{sec:est}

The goal is now to define estimation procedures for the three-parameters entering the definition of {\bf OSGRF} $X_{\theta_0,\alpha_0,H_0}$ and to study their statistical performance.
It is assumed, first, that $\theta_0$ is known and equal to $\theta_0 \equiv 0$ and estimation is devoted to parameters $\alpha_0$ and $H_0$.
In the second part, $\theta_0$ is unknown and needs to be estimated as well.
This is performed by applying the estimation of $\alpha_0$ and $H_0$ to a collection of rotated images.
It will be shown that the correct estimation angle is estimated when the estimation of the anisotropy coefficients reaches its minimum.

\subsection{Self-similarity and anisotropy parameters}
\label{sec:esta}

In this section, $\theta_0$ is assumed to be known and taken equal to $0$ for simplicity.  \\

\begin{figure}[t]
\begin{minipage}[t]{\linewidth}
\includegraphics[width=\linewidth]{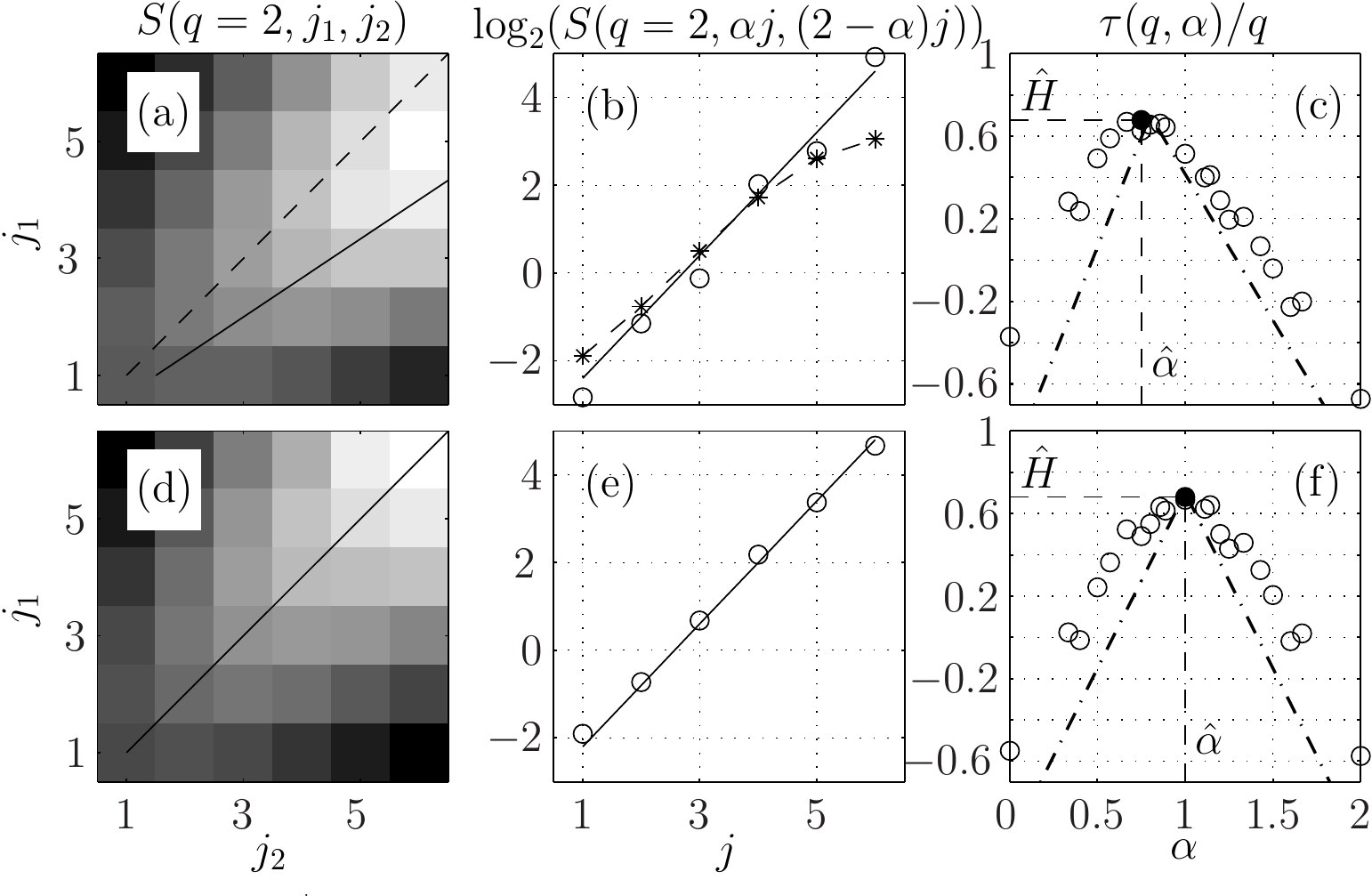}
\end{minipage}
\vspace{-.7cm}
\caption{\label{fig:estim} {\bf Illustrations of the estimation procedures.} For anisotropic, $\alpha_0 = 0.7$  (top row), and isotropic, $\alpha_0 = 1$ (bottom row), {\bf OSGRF} $X_{\theta_0,\alpha_0,H_0}$, with $\theta_0= 0, H_0=0.7$.
Left column: Structure functions $S(q,j_1,j_2)$. The solid line indicates the direction $\hat{\alpha}$ while the dashed lines corresponds to  $ \alpha = 1$ and hence the sole direction actually reachable with the coefficients of 2D-DWT.
Middle column: Estimation of $H(\alpha)$, based on $\tau(q=2,\alpha=\hat \alpha)$ obtained from a linear regression of $\log_2 S(q,\alpha j,(2-\alpha) j) $ versus $j$ ($\circ$).
Solid lines correspond to the theoretical $\tau(q=2,\alpha_0)$.
Stars (in (b)) correspond to the (biased) estimation of $H$ from $\tau(q=2,\alpha=1)$, i.e. by using only the 2D-DWT coefficients.
Right column: Plots of $\hat H_2(\alpha) = \tau(2,\alpha)/2$ versus $\alpha$.
The black dot shows the location of the maximum of $H_2(\alpha) = \tau(2,\alpha)/2$ thus yielding the estimated $\hat \alpha$ and $\hat H$.
In (f), as expected for an isotropic image, $\hat \alpha = 1$.
The mixed line corresponds to the theoretical values of $ \tau(2,\alpha)/2$ (cf. Eq. \ref{eq:estimalpha}).
 }
\end{figure}

\subsubsection{Estimation procedure}

By analogy to what has classically been done for the analysis of self-similarity, or scale invariance in general, (cf. e.g., \cite{abfrv02}), the space averages at joint scales $(j_1, j_2)$ (also referred to as structure functions) are used as estimators for the ensemble averages appearing in Eq. (\ref{eq:HWTq}) above:
\begin{equation}
\label{sec:tauqalpha}
 S(q,j_1,j_2) = \frac{1}{n_{j_1,j_2}}\sum_{(k_1,k_2)\in\mathbb{Z}^2} \vert d_X(j_1,j_2,k_1,k_2)\vert^q,
\end{equation}
where $n_{j_1,j_2}$ stands for the number of available coefficients jointly at scales $(2^{j_1},2^{j_2})$.

\noindent Let us further define $\tau_j(q,\alpha)$ as a function of the statistical order $q > 0$ and of the anisotropy parameter $\alpha$: 
\begin{equation}
\label{sec:tauqalpha}
\tau(q,\alpha)=\liminf_j\frac{-\alpha \log_2 (S(q,\alpha j,(2-\alpha)j)}{j }.
\end{equation}
In essence, Eq. (\ref{sec:tauqalpha}) amounts to assuming a power-law behavior of the structure functions with respect to scales, in the limit of fine scales $2^j \rightarrow +\infty$, along direction $\alpha$:
$$  S(q,\alpha j,(2-\alpha) j)  \simeq S_0(q) 2^{-j {\tau(q,\alpha)\over \alpha}}. $$

\noindent Eq. (\ref{eq:HWTq}) above indicates that, on average, and with the specific choice $(j_1, j_2) = (\alpha j,(2-\alpha)j)$:
$$ S(q,\alpha j,(2-\alpha) j) \approx 2^{j \frac{q}{\alpha}\left(1-(H_0+1)\max(\frac{\alpha}{\alpha_0},\frac{2-\alpha}{2-\alpha_0})  \right) }.  $$

\noindent Comparing these two last relations suggests that ${\tau(q,\alpha)\over \alpha} = \frac{q}{\alpha}\left(1-(H+1)\max(\frac{\alpha}{\alpha_0},\frac{2-\alpha}{2-\alpha_0})  \right)$, so that, for a given fixed $q$, the anisotropy parameter $\alpha_0$ can be estimated as: 
\begin{equation}
\label{eq:estimalpha}
\hat \alpha_{0,q}  =  \makebox{argmin}_\alpha \tau(q,\alpha),
\end{equation}
and that the self-similarity parameter $H$ can be estimated as: 
\begin{equation}
\label{eq:estimH}
\hat{H}_q  =  \tau(q,\hat{\alpha}_{0,q})/q.
\end{equation}

\subsubsection{Illustrations}
\label{sec:estillust}

The estimation procedure proposed here is sketched in Fig.~\ref{fig:estim}, for $q=2$, for an anisotropic (\ref{fig:estim}a-c) and an isotropic  (\ref{fig:estim}d-e)  {\bf OSGRF} $X_{\theta_0,\alpha_0,H_0}$.
It can be decomposed into three steps (for a given $q>0$).

{\bf Step 1:} The HWT coefficients $d_X(\underline{j},\underline{k})$ and corresponding structure functions $S(q,j_1,j_2)$ are computed.
Examples are shown in Fig.~\ref{fig:estim} (left column).

{\bf Step 2:} The surface $\log_2 S(q,j_1,j_2)$, seen as a function of the variables $j_1, j_2$ is interpolated  {(by nearest neighbor)} along the line $\alpha j_1+1=(2-\alpha) j_2+1$. Then, a non weighted least-square regression of  $\log_2 S(q,\alpha j,(2-\alpha) j)$ versus $\log_2 2^j = j$ is performed across all available scales, hence yielding an estimate of $\tau(q,\alpha)$, for each $\alpha$ and each $q$, as sketched in  Fig. \ref{fig:estim}-b and \ref{fig:estim}-e).

{\bf Step 3:} The estimated $\tau(q,\alpha)$ are plotted for a given $q$, as a function of $\alpha$, and its maximum yields the estimate $\hat{\alpha}$ of $\alpha_0$ (cf. Fig. \ref{fig:estim}-c and \ref{fig:estim}-f).
The estimation of the self--similarity parameter $H_0$ is further given by $\hat{H_q}=\tau(q,\hat{\alpha})/q$. \\
This procedure calls for the following comments.
First, Step~2 is performed for all accessible $\alpha$s, that is, for all values of $\alpha$, that connect at least two pairs of 
 dyadic scales, i.e.,  $(a_1=2^{j_1}, a_2=2^{j_2})$ with integers $(j_1,j_2)=[1, 2, \ldots, J]^2$. 
Therefore, the actual resolution of the values of $\alpha$  that can actually be used depends on the size $N \times N$ of the analyzed image, and hence so does the resolution of the estimate of the anisotropy parameter.
This discretized resolution can be observed in Fig. \ref{fig:test}.

Second, the structure functions $S(q,j,j)$ (hence for $\alpha=1$) are computed from HWT coefficients that actually corresponds to those of the 2D-DWT (cf. Fig.~\ref{fig:estim}-b, dashed line).
For isotropic fields, it is found that $\hat \alpha =1 $, and thus that the coefficients of the HWT that need to be actually used for the estimate of $H_0$ are those of the 2D-DWT.
Conversely, for anisotropic fields, basing the estimate of $H$ on $S(q,j,j)$ results into significant biases, as illustrated in Fig.~\ref{fig:estim}-b, dashed line and -e, where the estimate of $H_0$ for $\alpha \equiv 1$ significantly differs from that obtained for $\alpha = \hat \alpha$.
This illustrates the major benefits of replacing the 2D-DWT with the 2D-HWT.




\begin{figure}[t]
\begin{minipage}[t]{\linewidth}
\includegraphics[width=\linewidth]{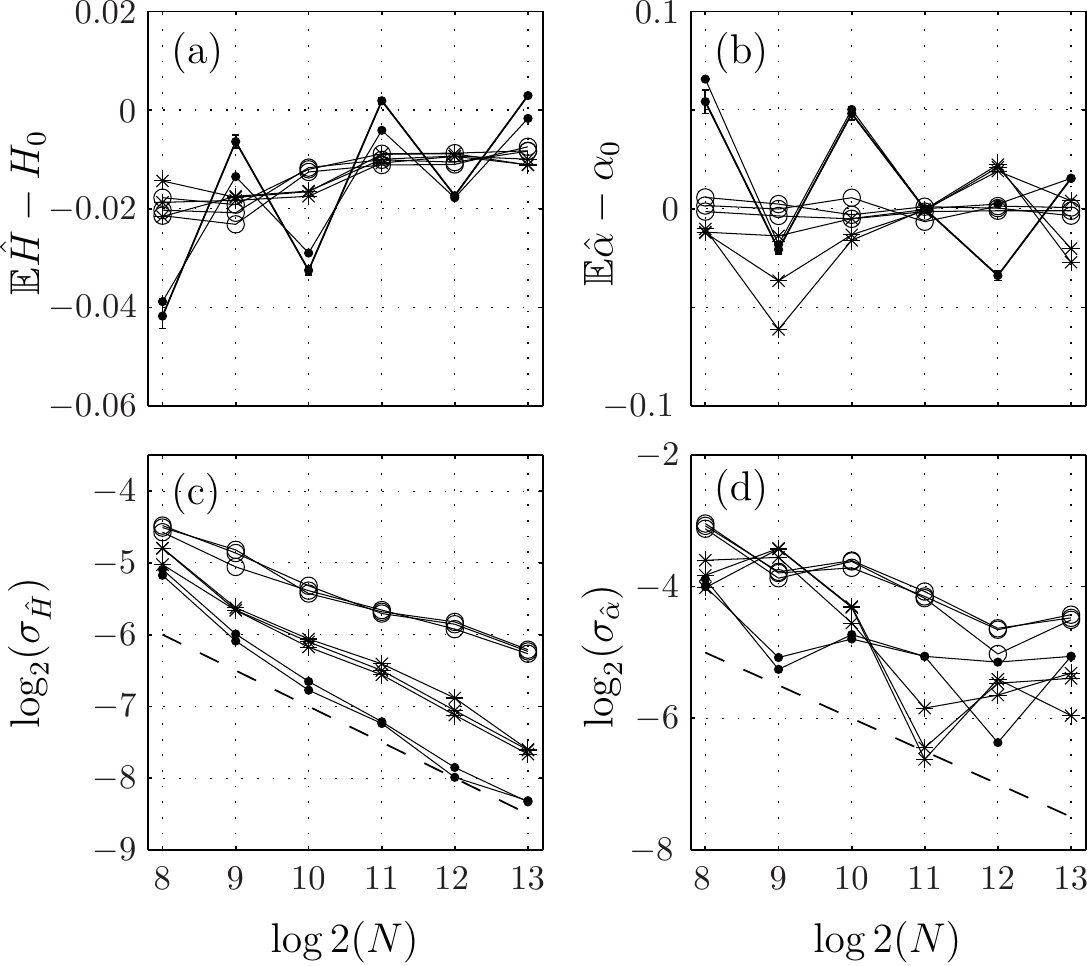}\\
\end{minipage}
\vspace{-.6cm}
\caption{\label{fig:perf} {\bf Estimation performance.} As functions of the sample size $N$ (image size $N \times N$), biases (top row) and standard deviation for $\hat H_{q=2} $ (left column) and $\hat \alpha_{q=2}$ (right column), obtained as average of estimation performed on $500$ realizations of  {\bf OSGRF} $X_{0,\alpha_0,H_0}$ with parameters $\alpha_0=1, H_0=0.7, 0.5$  and $0.3$ ($\circ$),
$\alpha_0=0.8, H_0=0.7, 0.5, 0.3$ ($\ast$) and $\alpha_0=0.6, H_0= 0.5, 0.3$ ({\large $\displaystyle \cdot$}).
Bottom row, dashed lines illustrates the expected $1/sqrt{N \times N}$ decrease of the standard deviations.
 }
\end{figure}

\subsubsection{Estimation Performance}

To complement the theoretical study reported above and to further assess the performance of the proposed estimation procedures, Monte-Carlo simulations are now performed further expanding on numerical investigations presented in  \cite{rcvja11}.
Biases and the standard deviations of $\hat \alpha_q$ and $\hat H_q$ are obtained from averages of estimates computed over $500$ independent realizations of {\bf OSGRF} $X_{0,\alpha_0,H_0}$, numerically produced by {\sc Matlab} routines designed by ourselves and available upon request.

Fig.~\ref{fig:perf} reports biases and  standard deviations as a function of (the $\log_2$ of)  the sample size $N$ (image size is $N\times N$), for various parameters $(\alpha_0,H_0)$. 
Fig.~\ref{fig:perf} essentially shows that the estimation performance both for $\hat \alpha $ and $\hat H$ does not depend on the actual $H_0$, a result that is highly reminiscent of the $1D$ case (cf. e.g., \cite{va01}).
However, dependences on the anisotropy parameter $\alpha_0$ do exist and are clearly visible on the standard deviation, which unexpectedly decrease with significant departures of isotropy.

Estimates are found to be asymptotically unbiased, as expected from theoretical analysis, and that standard deviations roughly decrease as $1/\sqrt(N\times N) = 1/N$, in agreement with the weak correlation property of the HWT coefficients.
Using other values of $q>0$ (ranging from $1$ to $5$) yields similar conclusions.

To conclude this section, let us put the emphasis on the fact that image sizes are varied from small  ($2^8 \times 2^8$) to (very) large ($2^{13} \times 2^{13}$).
This illustrates that both the synthesis and analysis procedures corresponding to the definition of {\bf OSGRF} and its analysis can be implemented efficiently and benefits from a remarkably low computational cost.

Estimation performance were reported here only for $q=2$, as it was found empirically that the use of other values of $q$ did not improve performance, as can be expected for Gaussian processes.

\begin{figure}[t]
\begin{minipage}[t]{\linewidth}
\includegraphics[width=\linewidth]{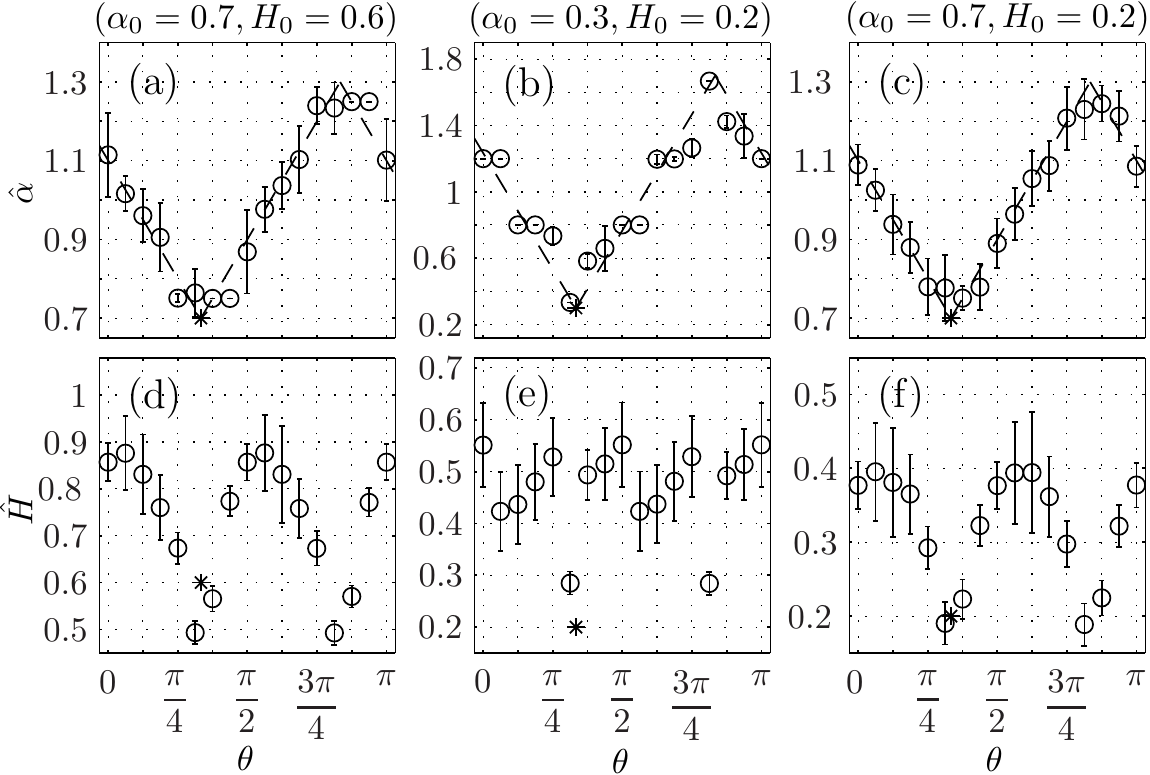}
\end{minipage}
\vspace{-.6cm}
\caption{\label{fig:results} {\bf Joint three parameter estimation procedure.} Estimations of $\hat{\alpha}$ (top row) and $\hat{H}$ (bottom row) for three differents fields with $\theta_0=\pi/3$ and
$(\alpha_0,H_0)=(0.7,0.6)$ ((a) and (d));  $(\alpha_0,H_0)=(0.3,0.2)$ ((b) and (e)); $(\alpha_0,H_0)=(0.7,0.2)$ ((c) and (f)).
The dashed line illustrates the expected theoretical behavior of $\hat{\alpha}$ as a function of $\theta$  and the 'circle' with confidence intervals to Monte-Carlo averages.
The estimation of $\hat{\theta}$ corresponds to the  location of the minimum of $\hat{\alpha}(\theta)$ and satisfactorily corresponds to $\theta_0=\pi/3$.
Final estimates for $\alpha_0$ and $H_0$ are obtained as $\hat{\alpha}^*=\hat{\alpha}(\hat{\theta})$ and $\hat{H}^*=\hat{H}(\hat{\theta})$ and thus show satisfactory agreement with the theoretical values, marked by $ \ast $.
Error bars correspond to $\sigma_{\hat H}$ (resp. $\sigma_{\hat \alpha}$).
}
\end{figure}

\subsection{Rotation parameter}
\label{sec:estb}

\subsubsection{Estimation procedure}

Let us now consider the case where, in addition to $H_0$ and $\alpha_0$, the rotation angle $\theta_0$ is unknown.
To estimate jointly the three unknown parameters, it is here proposed to apply the above procedure to estimate $H_0$ and $\alpha_0$ to a collection of rotated version of the original image, with rotation angles $\theta$.
The estimation of the anisotropy direction relies on the following observations, illustrated in Fig.~\ref{fig:results} (top row):
i) The estimate $\hat{\alpha}(\theta)$ is a $\pi$-periodic function~;
ii) it also has the symmetry $\hat{\alpha}(\theta_0+\theta)=\hat{\alpha}(\theta_0-\theta)$~;
iii) when $\theta=\theta_0$, $\hat{\alpha} \simeq \alpha$~;
iv) when  $\theta=\theta_0+\pi/2$, $\hat{\alpha}=2-\alpha$~;
v) and when  $\theta=\theta_0+\pi/4$, $\hat{\alpha}=1$.
Thus, the following joint estimation procedure for ($\theta_0, \alpha_0, H_0$) can be proposed:
\begin{eqnarray}
\hat \theta    & = & \makebox{argmin}_\theta \hat{\alpha}(\theta),\\
\hat{\alpha}^* & = & \hat{\alpha}(\hat{\theta})\\
 \hat{H}^*_q         & = & \hat{H}_q(\hat{\alpha}(\hat{\theta})).
\label{eq:estim2}
\end{eqnarray}
Because the minimum of  $ \hat{\alpha}(\theta)$ is (arbitrarily) picked, this procedure necessarily implies $\hat{\alpha}^* \leq 1$, there is thus a remaining indetermination whether the correct choice is $\hat{\alpha}^*$ or $2- \hat{\alpha}^*$ and therefore of $\pi/2$ in $\theta_0$.
As previously mentioned, this only amounts to exchanging the roles of the axis $x$ and $y$.
{For isotropic fields, $\hat \alpha(\theta)$ fluctuates around $\alpha=1$,  and no clear minimum (or maximum) is visible.
Furthermore,  $\hat \theta_{max}- \hat \theta_{min}$ differs from $\pi/2$.
When $\hat \theta_{max}-\hat \theta_{min}<\pi/4$, the field is thus declared isotropic and we set $\hat \theta =0$. }

To practically perform the rotation of $\theta$ on the image of analysis, a nearest neighbor interpolation is applied.
The procedure is totally automated and no human supervision is needed.

\subsubsection{Illustrations and performance}

To assess the performance of the joint three-parameter estimate procedure, Monte-Carlo numerical simulations are conducted, and biases and standard deviations are computed from average over 100 realizations of {\bf OSGRF} $X_{\theta_0,\alpha_0,H_0}$, for various choices of  $(\theta_0,\alpha_0,H_0)$, and with $q=2$.

Fig.~\ref{fig:results} shows, top row, that the estimation $\hat{\alpha}(\theta)$ clearly follows a piecewise linear variation along $\theta$ (modeled by the dashed line) and displays clear extrema for ${\theta} \approx \theta_0$ and ${\theta} \approx \theta_0+\pi/2$.
For  ${\theta} \approx \theta_0$,  $\hat{\alpha}(\hat{\theta})$ and $\hat{H}(\hat{\theta})$ (bottom line of Fig. \ref{fig:results}) provide satisfactory estimates of $\alpha_0$ and $H_0$.
For $\theta=\hat{\theta}+\pi/2$, the estimations are $2-\hat{\alpha}$ and $\hat H$. 
{Table \ref{table:test} displays the biases, standard deviations and Mean square errors for several isotropic and anisotropic  {\bf OSGRF} $X_{\theta_0,\alpha_0,H_0}$ fields.}
It can be observed that $\hat H$ shows more bias when a rotation of the original image is performed.
This is likely due to the interpolation procedure that smoothes out data and thus that \emph{distorts} self-similarity and thus scale invariance at the finest scales.
Better estimations for $H$ can be achieved by discarding a few of the finest scales from the linear regression, when image size permits.

\begin{figure}[t]
\begin{minipage}[t]{\linewidth}
\includegraphics[width=\linewidth]{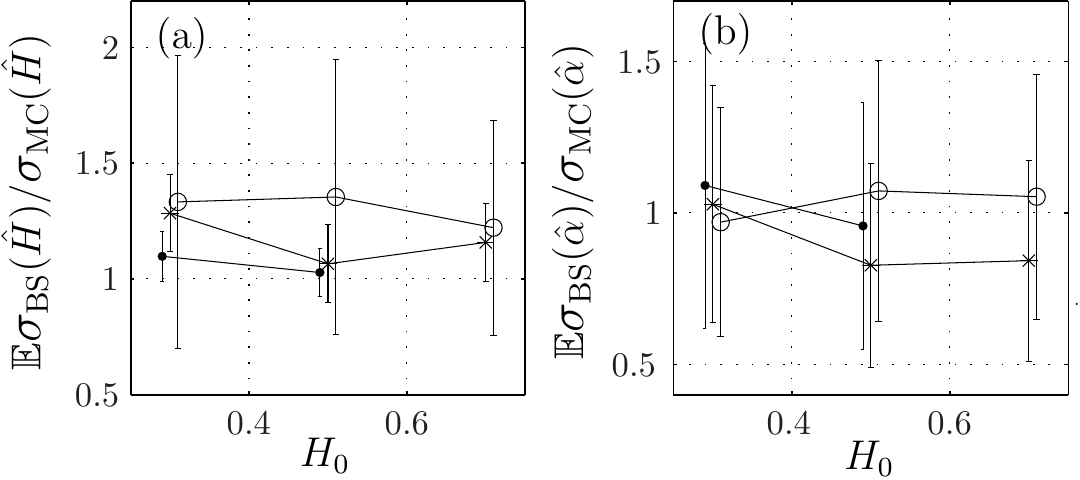}
\end{minipage}
\vspace{-.6cm}
\caption{\label{fig:BS} {\bf Bootstrap versus Monte Carlo Estimates of standard deviations.}  (a) Estimations of
$\displaystyle {\mathbb E} \sigma_{\mbox{\tiny BS}}(\hat H)/\sigma_{\mbox{\tiny MC}}(\hat H)$
as a function of $H_0$,
 for $q=2$, obtained from $R=100$ bootstraps applied to $100$ independent copies of {\bf OSGRF} $X_{0,\alpha_0,H_0}$ (image size $(2^{10},2^{10})$), with parameters $\alpha_0=1$
 ($\circ$), $\alpha_0=0.8$ ($\ast$)
and $\alpha_0=0.6$ ({\large $\displaystyle \cdot$}).
 (b) Estimations of  $\displaystyle {\mathbb E} \sigma_{\mbox{\tiny BS}}(\hat \alpha)/\sigma_{\mbox{\tiny MC}}(\hat \alpha)$ as a function of $H_0$,
for $\alpha_0=1$ ($\circ$), $\alpha_0=0.8$ ($\ast$)
and $\alpha_0=0.6$ ({\large $\displaystyle \cdot$}).
}
\end{figure}

\section{Bootstrap-based anisotropy test and confidence intervals}
\label{sec:test}

In applications, it is often of crucial importance to be able to test the isotropy assumption (i.e., whether $\alpha_0 = 1$ or not) for each single image independently.
This theoretically requires the knowledge of the distribution of  $\hat \alpha$.
Though it is found empirically Gaussian, the variance of the distribution remains unknown and, as suggested in Section~\ref{sec:esta} and Fig.~\ref{fig:perf}, it depends not only on the sample size $N$ but also on the unknown parameter $ \alpha_0$ itself.
Asymptotic Gaussian expansions for the calculations of the theoretical variance of  $\hat \alpha$, in the spirit of those proposed for fractional Brownian motion in e.g., \cite{va99}, have been observed to perform poorly (not reported here).
Instead, it is proposed to apply non parametric bootstrap procedure in the HWT coefficient domain, in the spirit of the procedures developed and assessed in \cite{WENDT:2007:E,W07,W08,W09}.
This procedure is detailed in the next section while the corresponding bootstrapped based isotropy test is defined and assessed in Section~\ref{sec:BSTest}.

\subsection{Bootstrap resampling schemes}
\label{sec:bs1}
In a nutshell, nonparametric bootstrap makes use of available samples, many times, by a drawing with replacement procedure, to yield an approximation of the  unknown  population distribution.
In turn, this estimated population distribution is used to construct confidence intervals or test (cf., e.g., \cite{Efron82} and \cite{Hall92}).

For the present work, following \cite{WENDT:2007:E}, the resampling procedure is applied in the HWT coefficient domain.
Because HWT coefficients do not consist of independent random variables, but possess a residual correlation, a time-block bootstrap procedure is used:
At each octave $\underline{j}$, block of size $l$ of HWT coefficients are drawn randomly with replacement.
This yields a set of bootstrapped HWT coefficients $d^{\ast}_X(\underline{j},\underline{k})$, from which bootstrap estimates $\hat \alpha^{\ast}$ and $\hat H^{\ast}$ of $\alpha_0$ and $H_0$, respectively, are obtained.

This procedure is repeated $R$ times, and the population distribution of $\hat \alpha$ and $\hat H$ are inferred from the boostrap estimates $\hat \alpha^{\ast,r}$ and $\hat H^{\ast,r}$, $r=1, \ldots, R$, notably variances can be estimated.

\subsection{Bootstrap-based estimates of variance}

It has been found empirically that $l$ need not depend on octave $\underline{j}$ and can be kept small. 
As documented in \cite{WENDT:2007:E},  $l$ is set to twice the size of the support of the mother wavelet (e.g., for a Daubechies3 wavelet used here $l=6$), as correlations amongst HWT coefficients is found to remain significant essentially over a space-scale controlled by the size of the wavelet support.

Fig.~\ref{fig:BS} compares the standard deviations of $\hat H$ (left) and $\hat \alpha $ obtained from 100 Monte Carlo simulations for anisotropic fields (of size $2^{10}\times 2^{10}$) against those obtained by  the bootstrap procedure (with $R=100$ for each of the $100$ Monte Carlo simulations).
Fig.~\ref{fig:BS} shows that the ratios $\displaystyle \sigma_{\mbox{\tiny BS}}(\hat H)/\sigma_{\mbox{\tiny MC}}(\hat H)$ and $\displaystyle \sigma_{\mbox{\tiny BS}}(\hat \alpha)/\sigma_{\mbox{\tiny MC}}(\hat \alpha)$ depend neither on  $\alpha_0$ nor on $H_0$ and remain close to $1$, with a slight overestimation  (from $10$ to $20 \%$) for the former and quasi perfect match for the latter.
Equivalent conclusions are drawn from different sample sizes $N$. 
These results indicate that the bootstrap estimates of the variances provide valuable approximations of the true variances of $\hat H$ and $\hat \alpha$.
Together with the Gaussian distribution empirical fact, this yields very satisfactory confidence intervals for $\hat H$ and $\hat \alpha$.

\begin{figure}[t]
\begin{minipage}[t]{\linewidth}
\includegraphics[width=\linewidth]{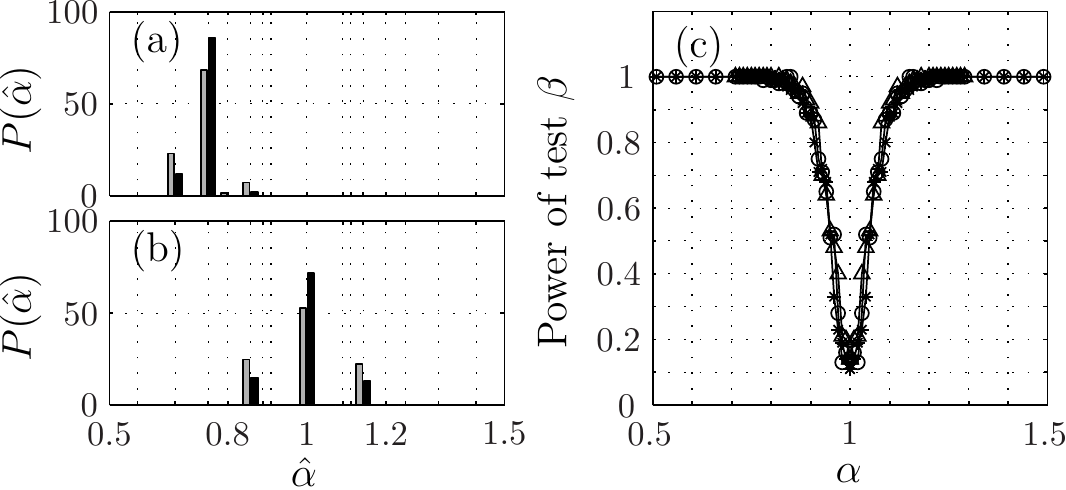}
\end{minipage}
\vspace{-.6cm}
\caption{\label{fig:test} {\bf Anisotropy test}. a) Histogram of $\hat\alpha_{\mbox{MC}}$ (light gray) and of $\hat\alpha_{\mbox{BS}}$ (black) for {\bf OSGRF} $X_{0,\alpha_0,H_0}$ (image size $(2^{10},2^{10})$), with $\alpha_0=0.75$ (a) and $\alpha_0=1$ (b).
Right plot (c) shows the rejection level of the test (with a significance level of $90\%$) obtained for $R=100$ bootstraps,  averaged on $100$ realizations of $X_{0,\alpha_0,H_0}$,  with $H_0=0.3$ ($\circ$), $0.5$ ($\vartriangle$) and $0.7$ ($\triangledown$).
}
\end{figure}

\subsection{Test procedure and performance}
\label{sec:BSTest}

\subsubsection{Test procedure}

To test isotropy in a given image, the null and alternative hypothesis respectively read:
\begin{equation}
{\cal H}_0 : \alpha_0-1=0, \; \mbox{ and } {\cal H}_A : \alpha_0-1 \neq 0.
\end{equation}

Let us assume first that $\theta_0 \equiv 0$.
The test procedure can be decomposed as follows:
\begin{itemize}
\item[-] Estimate $\hat \alpha$, as proposed in Section~\ref{sec:esta}.
\item[-] Apply the resampling scheme described in Section~\ref{sec:bs1} above to the HWT coefficients of $X_{\theta_0, \alpha_0, H_0} $ and construct the bootstrap distribution estimate of $\hat \alpha$ from the boostrap estimates $\hat \alpha^{\ast,r}, r= 1,\ldots R$.
\item[-] Set the test significance level $\delta $ for the test.
\item[-] Because when $\theta_0 \equiv 0$, there is no reason to decide a priori that the true $\alpha_0$ will depart from $1$ by being larger or smaller, a bilateral symmetric test is constructed.
Assuming a normal distribution for $\hat \alpha$, the bootstrap-based standard deviation
estimation $\sigma^*$  is used to construct the equi-tailed and symmetric acceptance region $ [-t_{\delta/2}\sigma^*,t_{\delta/2}\sigma^*]$, where $t_{\delta/2}$ denotes the $\delta/2$-th quantile of the zero-mean unit variance Gaussian distribution.
\item[-] Alternatively, the $p$-value of the test can be measured as the minimum between $P(\hat \alpha^{\ast}< \hat \alpha(\hat \theta)) $ and $1 - P(\hat \alpha^{\ast}>\hat \alpha(\hat \theta)) $, divided by $2$.
\end{itemize}

\subsubsection{Test performance}

To assess the validity and performance of the proposed test, it has been compared against Monte Carlo simulations,
based on $100$ independent copies of {\bf OSGRF} $X_{\theta_0=0,\alpha_0,H_0}$ with various parameter settings and for image size  $2^{10}\times 2^{10}$.
Fig.~\ref{fig:test}a) and ~\ref{fig:test}b)  compare the histograms of the estimates of $\alpha_0$ stemming from Monte Carlo simulations against those obtained from bootstrap estimates $\hat \alpha^*$, from a single realization, chosen arbitrarily, for anisotropic (a) and isotropic (b) fields.
For both cases, distributions are found to be in satisfactory agreement.
These figures also show that  $\hat \alpha$ can only take discretized values, because of the finite sample size of the image, as discussed in Section \ref{sec:estillust}.

In Fig.~\ref{fig:test}c) , the significance level of the test has been arbitrarily set to $\delta =0.9$ and the rejection level of the bootstrap test ($R=100$) has been computed as average over  $100$ independant Monte Carlo realizations of {\bf OSGRF} $X_{\theta_0=0,\alpha_0,H_0}$ for various parameter settings. 
When  $\alpha_0=1$, {\bf OSGRF} is isotropic and the rejection level $\beta$ is, as expected, found to satisfactorily reproduce the prescribed significance level $1-\delta =0.1$:
 $\hat\beta =0.13$, $0.15$ and $0.12$ respectively for $H_0=0.7, 0.5$ and $0.3$.
When $\alpha_0 \neq 1$, {\bf OSGRF} is anisotropic and the rejection level $\beta$ measures the power of the test.
Interestingly, it is found that the estimated power does not depend on  $H_0$, is symmetric for $\alpha_0$ above and below $1$ and mostly that it increases sharply when $\alpha_0$ departs from $1$.
This is thus indicating a strong potential to detect anisotropy even for small departure of $\alpha_0$ from $1$.
%
%
\begin{table}
\begin{center}
\begin{tabular}{||c|c|c|c|c||} \hline\hline
\multirow{2}*{$(\theta_0,\alpha_0,H_0)$} & $\langle \hat{\theta}\rangle-\theta_0$ & $\langle\hat{\alpha}\rangle-\alpha_0$ & $\langle\hat{H}\rangle-H_0$  &  \multirow{2}*{\% rej.} \\
 & (std,MSE) &  (std,MSE) &  (std,MSE) &  \\ \hline \hline
\multirow{2}*{$(\pi/3,0.7,0.6)$} & 0.01  & 0.00 & -0.10 & \multirow{2}*{100} \\
  &(0.03,0.00) & (0.04,0.00) & (0.02,0.01) & \\  \hline
\multirow{2}*{$(\pi/3,0.7,0.2)$} & 0.01  & 0.01 & -0.05 & \multirow{2}*{100} \\
  &(0.03,0.00) & (0.04,0.00) & (0.02,0.00) & \\  \hline
\multirow{2}*{$(\pi/3,0.3,0.2)$} & -0.01  & 0.01 & -0.09 & \multirow{2}*{100}\\
  &(0.02,0.00) & (0.05,0.00) & (0.09,0.02) & \\  \hline
\multirow{2}*{$(0,1,0.6)$} & 0.07  & 0.00 & -0.01 & \cellcolor[gray]{0.8}  \\
  &(0.31,0.01) & (0.08,0.01) & (0.03,0.00) & \cellcolor[gray]{0.8} \multirow{-2}*{8}\\  \hline
\multirow{2}*{$(0,1,0.2)$} & 0.08  & -0.01 & -0.01 & \cellcolor[gray]{0.8}\\
  &(0.25,0.01) & (0.08,0.01) & (0.03,0.00) & \cellcolor[gray]{0.8} \multirow{-2}*{11}\\  \hline \hline
\end{tabular}
\caption{\label{table:test} Biases, standard deviations and mean square errors obtained from $100$ independent copies of {\bf OSGRF} $X_{\theta_0,\alpha_0,H_0}$ (image size $(2^{10},2^{10})$). The right column reports the corresponding  rejection rate of the anisotropy test described in Section~\ref{sec:testtetha0not0}, with $R=100$ bootstrap surrogates. The significance level is set to $1- \delta = 10\%$.}
\end{center}
\end{table}

\subsubsection{Test procedure for  $\theta_0 \neq 0$}
\label{sec:testtetha0not0}

When $\theta_0$ is unknown and needs to be estimated, the procedure to test isotropy must be slightly amended, as follows:
\begin{itemize}
\item[-] Apply estimation procedure for $\theta_0, \alpha_0, H_0$ as in Section~\ref{sec:estb}.
\item[-] For $\hat \theta$, store the estimate $\hat \alpha(\hat \theta$) and the rotated field $\tilde{X}_{\hat \theta}$.
\item[-] Apply the resampling scheme described in Section~\ref{sec:bs1} above to the HWT coefficients of $\tilde{X}_{\hat \theta}$ and construct the bootstrap distribution estimate of $\hat \alpha$ from the boostrap estimates $\hat \alpha^{\ast,r}, r= 1,\ldots R$.
\item[-] Set the test significance level $\delta $ for the test.
\item[-] Because the estimated $\hat \alpha$ necessarily takes values in  $[0,1]$ a monolateral test must be constructed and the acceptance region is thus defined as:  $ [-t_{\delta}\sigma^*, 1]$.
\item[-] Alternatively, the $p$-value of the test can be computed as $P(\hat \alpha^{\ast}< \hat \alpha(\hat \theta))$.
\end{itemize}
{Table \ref{table:test} (right column) reports the rejection rates of the procedure applied to several anisotropic and isotropic  {\bf OSGRF} $X_{\theta_0,\alpha_0,H_0}$ fields of size $(2^{10},2^{10})$. For isotropic cases, the rejection rates matches closely the significance level, as expected.
For anisotropic fields, the power of the test is found very high as soon as $\alpha_0$ departs, even slightly, from $1$.}

\section{Other isotropic and anisotopic random fields}
\label{sec:othermodel}

So far, the analysis (estimation and test) procedures proposed here were applied only to the {\bf OSGRF} $X_{\theta_0,\alpha_0,H_0}$, defined in Section~\ref{sec:OSGRF}, and chosen as a convenient reference model, with three parameters accounting jointly for rotation, (an)isotropy and self-similarity.
However, one can naturally wonder whether the isotropy test described above would satisfactorily perform to detect anisotropy for other models, i.e., whether $\hat \alpha =1 $ or not.
In this section, a number of isotropic and non isotropic self-similar models commonly encountered in the image processing and statistics literature are used to test the level of generality of the approach proposed here.

\subsection{Random fields}

\subsubsection{Another OSGRF}

In \cite{br2011}, another interesting instance of {\bf OSGRF} has been explored. It is defined from Eq. (\ref{eq:modela}) with:
\begin{equation}
f(\underline{\xi})=(|\xi_1|^2+|\xi_2|^{2a})^{-\beta},
\label{eq:modelBE3}
\end{equation}
where $\beta=H_1+(1+1/a)/2$ and $a=H_2/H_1$ for $0<H_1<H_2<1$.
This process resembles {\bf OSGRF} $X_{\theta_0,\alpha_0, H_0}$, in Eq. (\ref{eq:model}), with $\alpha_0=2a/(1+a)$, $H_0=2aH_1/(1+a)$ and $\theta_0=0$.
It is thus anisotropic as soon as  $a \neq 1$.

\subsubsection{Extended Fractional Brownian Fields}
Another class of possibly anisotropic Gaussian field,  referred to as \emph{Extended Fractional Brownian Filed}, was first introduced in \cite{BE03}.
Its definition, $ {X_{f}(\underline{x})} =  \int_{\mathbb{R}^2} (e^{i\langle \underline{x}, \, \underline{\xi} \rangle} -1) f(\underline{\xi})^{1/2} d\widehat{W}(\underline{\xi})\, $,  relies on an admissible function $f$ of the form:
\begin{equation}
f(\underline{\xi})=|\underline{\xi}|^{-2h(arg(\underline{\xi}))-2},
\label{eq:modelBE2}
\end{equation}
where $arg(\underline{\xi})$ is the direction of the frequency $\underline{\xi}$ and $h$ an even measurable periodic function taking values in $(0,1)$.
Fractional Brownian field is a particular and isotropic case of {\bf EFBF}, where $h$ is a constant function, but {\bf EFBF}, is in general anisotropic when $h$ is not constant function. Strictly speaking, {\bf EFBF} is not exactly selfsimilar (except in cases where $h$ is  a constant function).
However,  {\bf EFBF} shows scale invariance properties that are empirically close to those of strictly selfsimilar fields.
Fig.~\ref{fig:modeleEFBF}a) shows a sample field of anisotropic {\bf EFBF}, with
\begin{equation}
h(arg(\underline{\xi}))=H_2\times(\cos(2\times arg(\underline{\xi}))+\epsilon)^2/(1+\epsilon)^2,
\end{equation}
where $\epsilon=1+2\sqrt{H_1/(H_2-H_1)}$. 
Function $h$ is $\pi$-periodic and takes values in $[H_1, H_2]$.
Fig. \ref{fig:modeleEFBF}c)  shows one sample-field, obtained with  parameters $H_1=0.2$ and $H_2=0.8$.

\begin{table}
\begin{center}
\begin{tabular}{||c||c|c|c||c|c|c||}  \hline\hline
$H_1$  & \multicolumn{3}{c||}{ $0.7$ } & \multicolumn{3}{c||}{ $0.4$ } \\
\cline{1-7}
$H_2$ & $0.5$ & $0.6$ & $0.7$ & $0.2$ & $0.3$ & $0.4$ \\ \hline \hline
OSGRF & 100 & 84 & \cellcolor[gray]{0.8}{15} & 100  & 99 & \cellcolor[gray]{0.8}{7} \\ \hline
EFBF & 42 & 31 & \cellcolor[gray]{0.8}{15} & 49 & 28 & \cellcolor[gray]{0.8}{10} \\ \hline
FBS & 89 & 78 & 58 & 92  & 87 & 61 \\ \hline \hline
\end{tabular}
\caption{\label{table:testa} {\bf Isotropie test: Rejection rates.} Obtained for three different classes of processes (from $R=100$ bootstraps on each of the $100$ $1024\times 1024$ realizations, significance level of $\delta = 90\%$).}
\end{center}
\end{table}

\subsubsection{Fractional Brownian Sheet }

Fractional Brownian Sheet ({\bf FBS}), introduced in~\cite{Kam96}, provides us with another class of (an)isotropic self-similar Gaussian field.
It can be defined through its harmonizable representation, for any $(H_1,H_2)$ in $(0,1)^2$ (see \cite{ALP02}) :
\begin{equation}\label{eq:defDBF}
B_{H_{1},H_{2}}(x) =\int_{\mathbb{R}^{2}}
\frac{(e^{i<x_{1},\xi_{1}>}-1)(e^{i<x_{2},\xi_{2}>}-1)}{|\xi_{1}|^{H_{1}+\frac{1}{2}}|\xi_{2}|^{H_{2}+\frac{1}{2}}}
d\widehat{W}_{\xi_{1},\xi_{2}},
\end{equation}
where $dW_{x_1,x_2}$ is a Brownian measure on $\mathbb{R}^{2}$ and $d\widehat{W}_{\xi_1,\xi_2}$ its Fourier transform.
{\bf FBS} is a Gaussian field with stationary rectangular increments, satisfying the following scaling property $\forall (a_1,a_2)\in(\mathbb{R}_+^*)^2$
\begin{equation}
\label{}
\{B_{H_1,H_2}(a_1 x_1,a_2 x_2)\}\overset{\mathcal{L}}{=}\{a_1^{H_1}a_2^{H_2}B_{H_1,H_2}(x_1,x_2)\}\;.
\end{equation}

\begin{figure}[t]
\begin{minipage}[t]{\linewidth}
\includegraphics[width=\linewidth]{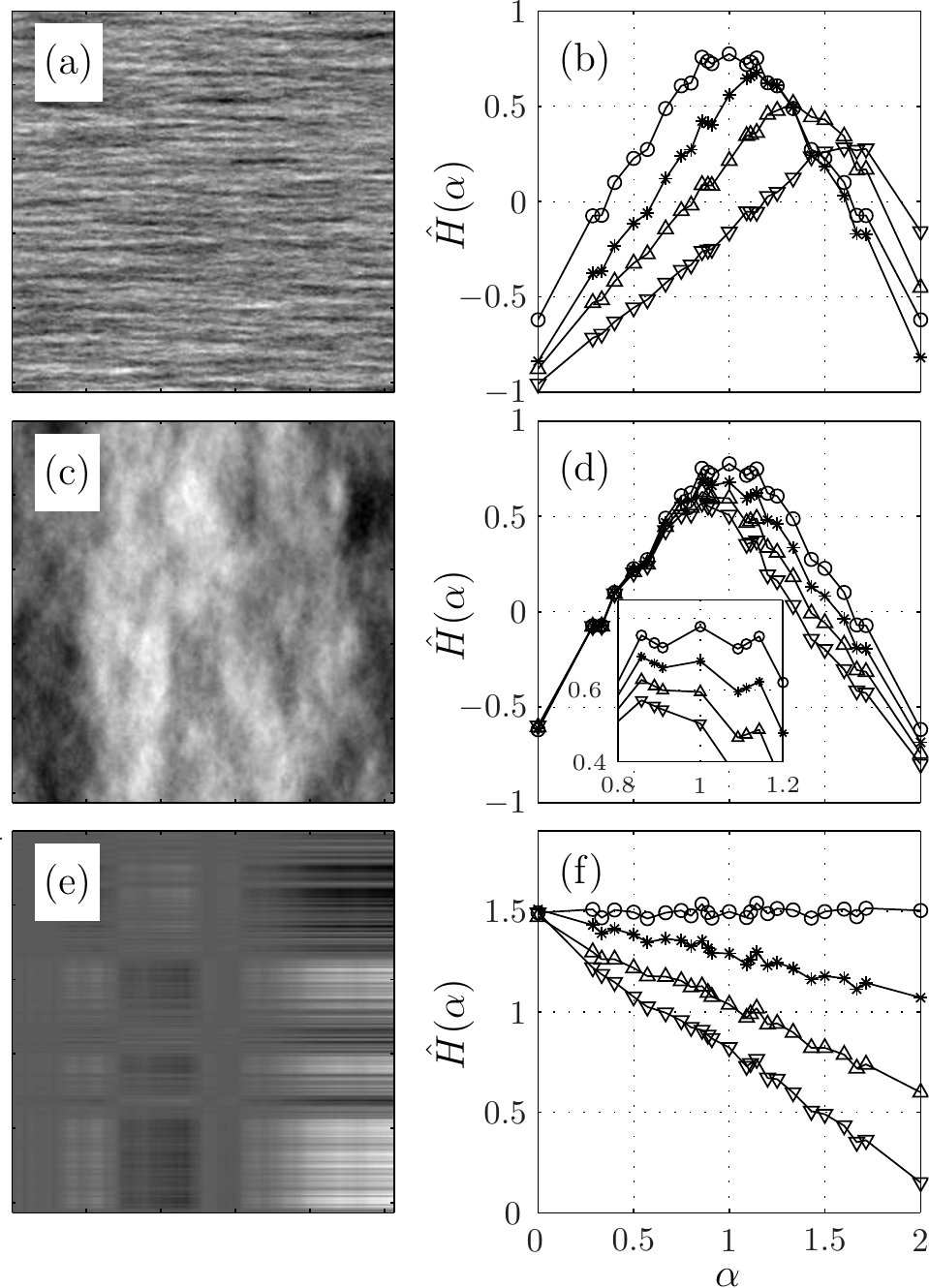}
\end{minipage}
\vspace{-.6cm}
\caption{\label{fig:modeleEFBF} {\bf Other (an)isotropic selfsimilar Gaussian fields.} Left column, sample fields with $(H_1,H_2)=(0.2,0.8)$ for {\bf OSGRF} (top), {\bf EFBF} (middle), {\bf FBS} (bottom).  Right column, $\hat H(\alpha)$ obtained for averages over $100$ realizations, with $H_2=0.8$ and $H_1=0.8$ ($\circ$), $0.6$ ($\star$), $0.4$ ($\vartriangle$) and $0.2$ ($\triangledown$).  $\hat H(\alpha)$ clearly shows a maximum for $\alpha \neq 1$ when fields are anisotropic.}
\end{figure}

\subsection{Testing anisotropy}

The estimation and test procedures described above were applied to these three classes of processes, for various setting of $[H_1, H_2]$.
Estimated function $\hat H(\alpha) $, averaged over $100$ realizations (size $2^{10}\times 2^{10}$), are reported in Fig. \ref{fig:modeleEFBF}, right column.
Isotropy rejection rates,  obtained from $R=100$ bootstrap surrogates for each of the $100$ realizations, are reported in Table \ref{table:testa}.

For the $3$ class of processes, when $H_1 \neq H_2$, it is observed that $\hat H(\alpha) $ has a maximum for $\alpha $ that clearly departs from $1$ and simultaneously that the isotropy rejection rates is far larger than the chosen $1 - \delta = 10\%$ significance level of the test.
This is the case even for as small discrepancies between $H_1$ and $H_2$, as $H_2-H_1= 0.2$.
These results clearly show that the proposed procedures clearly \emph{detect} anisotropy.

For {\bf EFBF}, it is reported in \cite{br2011} that the test anisotropy proposed therein failed to detect anisotropy (i.e., test reject in $0 \%$ of cases), when $H_1=0.5$ and $H_2=0.7$.
Trying as careful a comparison as possible, using the same model and parameter setting, it is found that the bootstrap test described in Section~\ref{sec:BSTest}, yields rejection of isotropy, with the $1-\delta = 10\% $ significance level, in $42 \%$ of cases, hence showing a much improved power (cf. Table \ref{table:testa}).

Conversely, for $H_1 = H_2$, it is observed for {\bf EFBF} and {\bf OSGRF} that $\hat H(\alpha) $ has a maximum for $\alpha =1$ and simultaneously that the isotropy rejection rates reproduces the targeted significance level, hence confirming that these processes are isotropic.
For {\bf FBS},  $\hat H(\alpha) $ remains flat for all $\alpha$s, while the rejection rates are higher than the  targeted significance level, this is thus questioning isotropy of {\bf FBS}, even when $H_1 = H_2$, a theoretically opened issue.

\begin{figure}[t]
\begin{minipage}[t]{\linewidth}
\includegraphics[width=\linewidth]{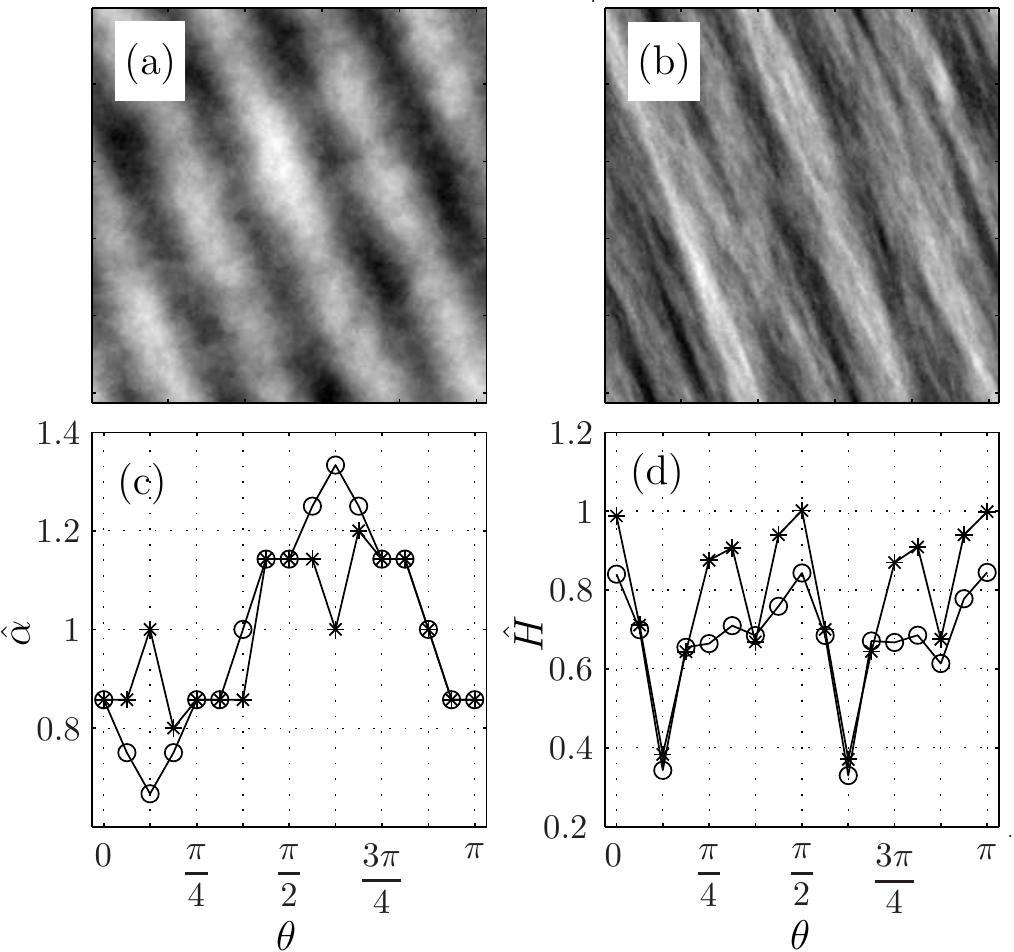}
\end{minipage}
\vspace{-.6cm}
\caption{\label{fig:sinus} (a) Sine wave of direction $\theta_0=\pi/8$ with an isotropic field $(H_0=0.5,\alpha_0=1)$; (b) anisotropic self-similar fields with $(H_0=0.5, \alpha_0=0.6,\theta_0=\pi/8)$. $\hat \alpha$ (c) and $\hat H$ (d) versus the angle analysis $\theta$. The symbols ($\circ$) represent the results obtained for the field in (a) and the ($\star$) for the field in (b).}
\end{figure}

\subsection{Anisotropic field with surimposed regular texture}

To finish, let us come back to the original issue disentangling self-similar with a true built-in anisotropy from isotropic selfsimilar processes to which an unrelated anisotropic texture is additively superimposed.
To address this issue, let us compare a truly isotropic {\bf OSGRF} $X_{\theta_0=0, \alpha_0=1, H_0=0.5}$, as defined in Eq. (\ref{eq:model}), to which a sine waveform trend,  with orientation $\theta_0=\pi/8$ is additively superimposed (Fig.~\ref{fig:sinus}a) to a truly anisotropic {\bf OSGRF} $X_{\theta_0=\pi/8, \alpha_0=0.6, H_0=0.5}$.
The estimation and test procedures described above are applied to $100$ realizations of both processes, and $\hat \alpha(\theta)$ and $H(\theta)$ are displayed in Fig.~\ref{fig:sinus}c and d, respectively.
For the truly anisotropic field ($\circ$), $\hat \alpha(\theta)$ displays a clear minimum for $\hat \theta= \theta_0$, with estimated anisotropy ($\hat \alpha=\alpha(\hat \theta)=0.66$) and selfsimilarity ($\hat H(\hat \theta)= 0.36$) parameters in close agreement with the true ones ($\circ$ in \ref{fig:sinus}d).
This is thus clearly validating anisotropy.
For the isotropic field, to which the directional sine wave trend has been added, $\alpha(\theta) $ shows no clear minimum and instead a rather constant behavior in $\theta$ is observed, thus leading to conclude that  anisotropy,  clearly visible on the sample field, is superimposed on rather than built-in self-similarity.

This example leads us to conclude that the  procedure  proposed in the present contribution provides practitioners with a reliable tool to analyze self-similarity in presence of anisotropy and enables them to clearly disentangle built-in anisotropy from independent and superimposed added anistropic trends.

\section{Conclusions and perspectives}
\label{sec:discussion}

The present contribution aimed at studying images or fields where self-similarity is potentially tied to anisotropy.
Replacing the standard 2D-DWT with the HWT, thus permitting to use different dilation factors along horizontal  and vertical directions, enabled us first to estimate the rotation and anisotropy parameters.
In turn, this permitted a correct estimation of the self-similarity parameter along the estimated anisotropy direction.
This direction selection would not be permitted by the use of the sole 2D-DWT coefficients and thus constitutes the major benefits of the use of the HWT, and therefore the key feature of the present contribution.

Additionally, bootstrap based procedures, performed in the HWT coefficient domain, furnish confidence intervals for the estimates and an isotropy test, that can be applied to a single image.

Though studied in depth for a specific Gaussian self-similar model, the proposed analysis is shown to enable the detection of  anisotropy for a large variety of classes of  Gaussian self-similar processes.
Also, true built-in anisotropy is clearly discriminated from isotropy to which an anisotropic trend is added. \\

Extensions of the applicability of the present method or further developments geared towards the analysis of more general classes of processes modeling. Textures with scale invariance, that are not necessarily exactly self-similar and that may, weakly or significantly, depart from Gaussian distributions, are under current investigations.
Notably, this study paves the way toward the far more difficult topic of multifractal analysis and formalim in presence of anisotropy, to which future efforts are devoted.

{\sc Matlab} routines, designed by ourselves, implementing field synthesis and parameter estimation and test will be made publicly available at the time of publication.



\bibliographystyle{IEEEtran}
\bibliography{anisotopy_new}

\begin{thebibliography}{10}
\providecommand{\url}[1]{#1}
\csname url@samestyle\endcsname
\providecommand{\newblock}{\relax}
\providecommand{\bibinfo}[2]{#2}
\providecommand{\BIBentrySTDinterwordspacing}{\spaceskip=0pt\relax}
\providecommand{\BIBentryALTinterwordstretchfactor}{4}
\providecommand{\BIBentryALTinterwordspacing}{\spaceskip=\fontdimen2\font plus
\BIBentryALTinterwordstretchfactor\fontdimen3\font minus
  \fontdimen4\font\relax}
\providecommand{\BIBforeignlanguage}[2]{{%
\expandafter\ifx\csname l@#1\endcsname\relax
\typeout{** WARNING: IEEEtran.bst: No hyphenation pattern has been}%
\typeout{** loaded for the language `#1'. Using the pattern for}%
\typeout{** the default language instead.}%
\else
\language=\csname l@#1\endcsname
\fi
#2}}
\providecommand{\BIBdecl}{\relax}
\BIBdecl

\bibitem{RAD2000}
S.~G. Roux, A.~Arneodo, and N.~Decoster, ``A wavelet-based method for
  multifractal image analysis. iii. applications to high-resolution satellite
  images of cloud structure,'' \emph{European Physical Journal B}, vol.~15,
  no.~4, pp. 765--786, Jun. 2000.

\bibitem{f97}
P.~Frankhauser, ``L'approche fractale : un nouvel outil dans l'analyse spatiale
  des agglomerations urbaines,'' \emph{Population}, vol.~4, pp. 1005--1040,
  1997.

\bibitem{Lundahl:1986}
T.~Lundahl, W.~J. Ohley, S.~M. Kay, and R.~Siffert, ``Fractional brownian
  motion: A maximum likelihood estimator and its application to image
  textures,'' \emph{IEEE Trans. Medical Imaging}, vol.~5, no.~3, 1986.

\bibitem{aK2001}
P.~Kestener, J.-M. Lina, P.~Saint-Jean, and A.~Arneodo, ``Wavelet-based
  multifractal formalism to assist in diagnosis in digitized mammograms,''
  \emph{Image Anal. Stereol.}, vol.~20, pp. 169--174, 2001.

\bibitem{Rachidi::2008}
M.~Rachidi, F.~Richard, H.~Bierme, C.~Roux, P.~Fardellone, E.~Lespessailles,
  C.~Chappard, and C.~Benhamou, ``Osteoporosis risk assessment: A composite
  index combining clinical risk factors and biophysical parameters,''
  \emph{Journal of Bone and Mineral Research}, vol.~23, pp. S112--S112, Sep.
  2008.

\bibitem{Richard::2010}
F.~Richard and H.~Bierme, ``Statistical tests of anisotropy for fractional
  brownian textures. application to full-field digital mammography,''
  \emph{Journal of Mathematical Imaging and Vision}, vol.~36, no.~3, pp.
  227--240, Mar. 2010.

\bibitem{B2010}
M.~Bergounioux and L.~Piffet, ``A second-order model for image denoising,''
  \emph{Set Valued and Variational Analysis}, vol.~18, no. 3-4, pp. 277--306,
  2010.

\bibitem{SL87}
D.~Schertzer and S.~Lovejoy, ``Physically based rain and cloud modeling by
  anisotropic, multiplicative turbulent cascades,'' \emph{J. Geophys. Res.},
  vol.~92, pp. 9693--9714, 1987.

\bibitem{ABRY:2012:A}
P.~Abry, J.~S.2, and H.~Wendt, ``When {V}an {G}ogh meets {M}andelbrot:
  Multifractal classification of painting textures,'' \emph{Signal Processing},
  2012, to appear.

\bibitem{U1995}
M.~Unser, ``Texture classification and segmentation using wavelet frames.''
  \emph{IEEE Transactions on Image Processing}, vol.~4, no.~11, pp. 1549--1560,
  1995.

\bibitem{N02}
M.~Nielsen, L.~K. Hansen, P.~Johansen, and J.~Sporring, ``Guest editorial:
  Special issue on statistics of shapes and textures,'' \emph{Journal of
  Mathematical Imaging and Vision}, vol.~17, no.~2, pp. 87--87, Sep. 2002.

\bibitem{D2002}
M.~Do and M.~Vetterli, ``Wavelet-based texture retrieval using generalized
  gaussian density and kullback-leibler distance.'' \emph{IEEE Transactions On
  Image Processing}, vol.~11, no.~2, pp. 146--158, 2002.

\bibitem{C05}
M.~Chantler and L.~Van~Gool, ``Special issue on "texture analysis and
  synthesis",'' \emph{International Journal of Computer Vision}, vol.~62, no.
  1-2, pp. 5--5, Apr. 2005.

\bibitem{vsv99}
G.~Van~de Wouwer, P.~Scheunders, and D.~Van~Dyck, ``Statistical texture
  characterization from discrete wavelet representations.'' \emph{IEEE
  Transactions On Image Processing}, vol.~8, no.~4, pp. 592--598, 1999.

\bibitem{P84}
S.~Peleg, J.~Naor, R.~Hartley, and D.~Avnir, ``Multiple resolution texture
  analysis and classification,'' \emph{IEEE Transactions On Pattern Analysis
  and Machine Intelligence}, vol.~6, no.~4, pp. 518--523, 1984.

\bibitem{K99}
L.~M. Kaplan, ``Extended fractal analysis for texture classification and
  segmentation,'' \emph{IEEE Transactions On Image Processing}, vol.~8, no.~11,
  pp. 1572--1585, Nov. 1999.

\bibitem{Pesquet-Popescu::2002}
B.~Pesquet-Popescu and J.~L. Vehel, ``Stochastic fractal models for image
  processing,'' \emph{IEEE Signal Processing Magazine}, vol.~19, no.~5, pp.
  48--62, Sep. 2002.

\bibitem{BE03}
A.~Bonami and A.~Estrade, ``Anisotropic analysis of some gaussian models,''
  \emph{The Journal of Fourier Analysis and Applications}, vol.~9, no.~3, pp.
  215--236, 2003.

\bibitem{BMS07}
H.~Bierm{\'e}, M.~Meerschaert, and H.~Scheffler, ``Operator scaling stable
  random fields.'' \emph{Stoch. Proc. Appl.}, vol. 117, no.~3, pp. 312--332,
  2009.

\bibitem{R92}
A.~R. Rao and R.~C. Jain, ``Computerized flow field analysis - oriented texture
  fields,'' \emph{IEEE Transactions On Pattern Analysis and Machine
  Intelligence}, vol.~14, no.~7, pp. 693--709, Jul. 1992.

\bibitem{S94}
C.~F. Shu and R.~C. Jain, ``Vector field analysis for oriented patterns,''
  \emph{IEEE Transactions On Pattern Analysis and Machine Intelligence},
  vol.~16, no.~9, pp. 946--950, Sep. 1994.

\bibitem{rcvja11}
S.~Roux, M.~Clausel, B.~Vedel, S.~Jaffard, and P.~Abry, ``Transform\'e
  hyperbolique en ondelettes 2d pour la caract\'erisation d'images
  autosimilaires anisotropes,'' in \emph{XXIII colloque sur le Traitement du
  Signal et des Images GRETSI, Bordeaux, France}, 5--8 Sep. 2011.

\bibitem{DeVore1998}
R.~A. DeVore, S.~V. Konyagin, and V.~N. Temlyakov, ``Hyperbolic wavelet
  approximation,'' \emph{Constructive Approximation}, vol.~14, pp. 1--26, 1998.

\bibitem{Yu96}
T.~Yu, A.~AStoschek, and D.~L. Donoho, ``Translation- and direction-invariant
  denoising of 2d and 3d images: experience and algorithms,'' in \emph{Proc.
  SPIE 2825, Wavelet Applications in Signal and Image Processing IV}, 1996, p.
  608.

\bibitem{Ro1999}
C.~P. Rosiene and T.~Q. Nguyen, ``Tensor-product wavelet vs. {M}allat
  decomposition: a comparative analysis,'' in \emph{ISCAS'99}, vol.~3, Jul.
  1999, pp. 431--434.

\bibitem{Zav07}
V.~Zavadsky, ``Image approximation by rectangular wavelet transform,''
  \emph{Journal of Mathematical Imaging and Vision,}, vol.~27, no.~2, pp.
  129--138, 2007.

\bibitem{S05}
I.~W. Selesnick, R.~G. Baraniuk, and N.~G. Kingsbury, ``The dual-tree complex
  wavelet transform,'' \emph{Ieee Signal Processing Magazine}, vol.~22, no.~6,
  pp. 123--151, Nov. 2005.

\bibitem{WENDT:2007:E}
H.~Wendt, P.~Abry, and S.~Jaffard, ``Bootstrap for empirical multifractal
  analysis,'' \emph{IEEE Signal Processing Mag.}, vol.~24, no.~4, pp. 38--48,
  2007.

\bibitem{Mallat1998}
S.~Mallat, \emph{A Wavelet Tour of Signal Processing}.\hskip 1em plus 0.5em
  minus 0.4em\relax San Diego, CA: Academic Press, 1998.

\bibitem{Trie06}
H.~Triebel, \emph{Theory of functions spaces {I}{I}{I}}.\hskip 1em plus 0.5em
  minus 0.4em\relax Birkhauser Verlag, 2006.

\bibitem{BMA93}
E.~Bacry, J.~Muzy, and A.~Arneodo, ``Singularity spectrum of fractal signals
  from wavelet analysis - exact results,'' \emph{Journal Of Statistical
  Physics}, vol.~70, no. 3-4, pp. 635--674, FEB 1993.

\bibitem{daubechies92}
I.~Daubechies, \emph{Ten {L}ectures on {W}avelets}.\hskip 1em plus 0.5em minus
  0.4em\relax Philadelphia: SIAM, 1992.

\bibitem{CV10b}
M.~Clausel and B.~Vedel, ``An optimality results about sample path properties
  of operator scaling gaussian random fields,'' \emph{arXiv:1302.0818}, 2010.

\bibitem{ACJRV2012}
P.~Abry, M.~Clausel, S.~Jaffard, S.~Roux, and B.~Vedel, ``Hyperbolic wavelet
  transform: an efficient tool for multifractal analysis of anisotropic
  textures,'' \emph{arXiv:1210.1944}, Oct. 2012.

\bibitem{CV10a}
M.~Clausel and B.~Vedel, ``Explicit constructions of operator scaling
  self-similar random gaussian fields,'' \emph{Fractals}, vol.~19, no.~1, pp.
  101--111, 2011.

\bibitem{abfrv02}
P.~Abry, R.~Baraniuk, P.~Flandrin, R.~Riedi, and D.~Veitch, ``Multiscale
  network traffic analysis, modeling, and inference using wavelets,
  multifractals, and cascades,'' \emph{IEEE Signal Processing Magazine},
  vol.~3, no.~19, pp. 28--46, May 2002.

\bibitem{va01}
D.~Veitch and P.~Abry, ``A statistical test for the time constancy of scaling
  exponents,'' \emph{IEEE Trans. on Sig. Proc.}, vol.~49, no.~10, pp.
  2325--2334, 2001.

\bibitem{va99}
------, ``A wavelet based joint estimator of the parameters of long-range
  dependence,'' \emph{IEEE Transactions on Information Theory special issue on
  "Multiscale Statistical Signal Analysis and its Applications"}, vol.~45,
  no.~3, pp. 878--897, April 1999.

\bibitem{W07}
H.~Wendt and P.~Abry, ``Multifractality tests using bootstrapped wavelet
  leaders,'' \emph{IEEE Transactions On Signal Processing}, vol.~55, no.~10,
  pp. 4811--4820, Oct. 2007.

\bibitem{W08}
------, ``Bootstrap tests for the time constancy of multifractal attributes,''
  \emph{Proceedings of IEEE International Conference On Acoustics, Speech and
  Signal Processing,}, vol. 1-12, pp. 3465--3468, 2008.

\bibitem{W09}
H.~Wendt, S.~Roux, S.~Jaffard, and P.~Abry, ``Wavelet leaders and bootstrap for
  multifractal analysis of images,'' \emph{Signal Processing}, vol.~89, no.~6,
  pp. 1100--1114, Jun. 2009.

\bibitem{Efron82}
B.~Efron, \emph{The Jackknife, the Bootstrap and Other Resampling Plans},
  S.~I.~A. MEditors, Ed.\hskip 1em plus 0.5em minus 0.4em\relax Society for
  Industrial and Applied Mathematics, 1982, vol.~38, no.~38.

\bibitem{Hall92}
P.~Hall, \emph{The Bootstrap and Edgeworth Expansion}.\hskip 1em plus 0.5em
  minus 0.4em\relax Springer Verlag, 1992.

\bibitem{br2011}
H.~Bierm{\'e} and F.~Richard, ``Analysis of texture anisotropy based on some
  gaussian fields with spectral density,'' in \emph{Mathematical Image
  Processing}, ser. Springer Proceedings in Mathematics, M.~Bergounioux, Ed.,
  vol.~5.\hskip 1em plus 0.5em minus 0.4em\relax Springer, 2011, pp. 59--73.

\bibitem{Kam96}
A.~Kamont, ``On the fractional anisotropic wiener field,'' \emph{Probability
  and mathematical statistics}, vol.~16, no.~1, pp. 85--98, 1996.

\bibitem{ALP02}
A.~Ayache, S.~L{\'e}ger, and M.~Pontier, ``Drap brownien fractionnaire.''
  \emph{Pot. Anal.}, vol.~17, pp. 31--43, 2002.

\end{thebibliography}

\end{document}